\documentclass[a4paper,11pt]{article}

\usepackage{jheppub} 
\usepackage{physics}
\usepackage{subfigure}

\preprint{KUNS-3071}

\title{\boldmath 
Gluon scattering amplitudes with instantons and minimal surfaces with topology change
}

\author[a]{Koji Hashimoto,}
\author[a]{Koichi Kyo,}
\author[b]{Masaki Murata,}
\author[b]{Gakuto Ogiwara,}
\author[a]{Norihiro Tanahashi}

\affiliation[a]{Department of Physics, Kyoto University, Kyoto 606-8502, Japan}
\affiliation[b]{Department of Information Systems, Saitama Institute of Technology, Saitama
369-0293, Japan}

\emailAdd{koji@scphys.kyoto-u.ac.jp}
\emailAdd{kyo.koichi@gauge.scphys.kyoto-u.ac.jp}
\emailAdd{m.murata@sit.ac.jp}
\emailAdd{i5001hrr@sit.ac.jp}
\emailAdd{tanahashi@gauge.scphys.kyoto-u.ac.jp}

\abstract{
We study the instanton effect on the gluon scattering amplitudes at strong coupling and large $N$ for the ${\cal N}=4$ supersymmetric Yang-Mills theory. According to Alday and Maldacena, the gluon scattering amplitude corresponds holographically to the area of a worldsheet minimal surface in the T-dual AdS${}_5$ geometry. The Yang-Mills instanton introduces an instanton D-brane in the geometry, with which a particular boundary condition for the minimal surface is imposed. We show that the minimal surface undergoes a topology change depending on the size and the gluon momenta, and that the instanton amplitude exhibits a characteristic dependence on gluon momenta. More specifically, we find that when the fixed instanton size modulus $\rho$ is larger than ${\cal O}(\sqrt{\lambda}/E)$ where $\lambda$ is the 't Hooft coupling and $E$ is the typical momentum of the scattering gluon, due to the topology change of the worldsheet minimal surface, the instanton amplitude is exponentially enhanced as $\exp(\rho E)$.
}

\begin{document}

\maketitle
\flushbottom

\section{Introduction}
\label{sec:intro}

Yang-Mills instantons are one of the most important non-perturbative ingredients in quantum field theories. As the Yang-Mills theories exhibit the confinement and other related strong-coupling effects as emergent phenomena, to find out the origin of those phenomena in an understandable manner is one of the grand challenges in the research of quantum field theories. As the instanton contributions to any correlation function and thus any scattering amplitude would come as $\exp[-1/g^2]$ where $g$ is the Yang-Mills coupling constant, when we treat the regime of a strong coupling, all instanton amplitudes, the scattering amplitudes in the presence of the instantons, participate equally in the phenomena, which would contribute significantly to the emergent phenomena mentioned above. Thus, controllable setups to analyze the instanton amplitudes are welcome, to discover a possible understandable explanation of the emergent phenomena of strongly coupled gauge theories.

Although the classical instantons and their quantization via the moduli space are well-studied, how they play their role in scattering amplitudes yet waits its exploration. In QCD, instanton contribution to the QCD scattering amplitudes (see for example \cite{Khoze:2019jta}\footnote{For earlier calculations of the instanton amplitudes, see \cite{Khoze:1991mx,mueller1991first, mueller1991higher, Shuryak:2000df, Kharzeev:2000ef}.}) suggests that instanton-induced processes may provide soft particle production at collider experiments. However, the instanton expansion makes better sense at weak coupling, and it would be desired to find and check a universal feature of the instanton scattering amplitude at strong coupling.

In this paper, we resort to a holographic method to find a universal behavior of the instanton amplitudes. According to Alday and Maldacena \cite{Alday:2007hr}, gluon scattering amplitudes in ${\cal N}=4$ supersymmetric Yang-Mills theory (SYM) at strong coupling and large $N$ limit can be calculated holographically, under the AdS/CFT correspondence \cite{Maldacena:1997re}.\footnote{In the holographic dictionary, normally the correlators of gauge-invariant operators are considered, while the dictionary for the scattering amplitudes of gluons is one rare example of extracting the information of non-gauge-invariant gluon operators. This may sound a little contradictory, but it would be resolved if one is reminded of that the Wilson loop has a gravity dual and in fact Wilson loop is expanded to a product of gluon operators, which may be considered to be leading to the scattering amplitudes of gluons. In fact, as we review briefly below, the dictionary for the gluon scattering amplitudes is closely related to that for the Wilson loop, as anticipated in \cite{Alday:2007he,Drummond:2007aua}.}
We introduce Yang-Mills instantons to this holographic set-up and calculate the gluon scattering amplitudes in the presence of multiple Yang-Mills instantons.\footnote{The instanton corrections to the correlators of gauge invariant operators in a certain limit in the ${\cal N}=4$ SYM vanish \cite{Bianchi:1998nk}. Our target is the gluon scattering amplitudes for which the argument does not apply.}

Alday and Maldacena found that the $n$-gluon scattering amplitude is proportional to $\exp[-S_{\rm NG}]$ where $S_{\rm NG}$ is the Nambu-Goto action of a string worldsheet in the so-called T-dual AdS geometry. At its AdS boundary, the worldsheet ends on a null $n$-polygon whose edges are the momenta of the gluons. The topology of this worldsheet is a disk. They succeeded in evaluating the gluon 4-point amplitude analytically. 
What needs emphasis is that, when the number of gluons increases, the evaluation of the worldsheet area becomes more difficult and numerical evaluation is necessary. The authors of \cite{Dobashi:2008ia,Dobashi:2009sj} used a numerical method of mesh discretization of the worldsheet to calculate the 6-point amplitudes numerically.

On the other hand, instantons in Yang-Mills theories are holographically dual to D-instantons in the original AdS \cite{Banks:1998nr}, which actually served as an evidence of the AdS/CFT correspondence in early days --- the Moduli space metric of a supersymmetric Yang-Mills instanton in fact is that of the AdS${}_5$ spacetime \cite{Bianchi:1998nk,Dorey:1998qh,Dorey:1999pd}. Therefore, in the gravity side, the instanton scattering amplitude should correspond to the worldsheet in the presence of the D-instantons.

In this paper, 
following Alday and Maldacena, we perform a T-duality to go to the T-dual AdS. Then the D-instanton is T-dualized to a D3-brane which is parallel to the
boundary of the T-dual AdS. This D3-brane actually ``cuts" the worldsheet to make a hole, since the worldsheet can end on D-branes. The topology of the worldsheet can change from the disk to a cylinder. The area of the worldsheet which ends on the D3-brane provides the effect of the instantons against the gluon scattering amplitudes in the ${\cal N}=4$ SYM.

To evaluate the worldsheet area to get the instanton amplitudes holographically, we need to determine the worldsheet configuration which has to satisfy a set of boundary conditions: at the T-dual AdS boundary it should end on the null polygon, while at the D3-brane which is T-dual to the D-instanton it should satisfy the Neumann boundary condition. 
In this paper, based on the knowledge of the generic behavior of the minimal area worldsheet in holographic spacetime, we perform a model analysis of the worldsheet to obtain a quantitative behavior of the gluon scattering amplitudes as a function of the instanton size. We find that, for a fixed size $\rho$ of the instanton, the worldsheet undergoes a topology change when the gluon energy $E$ increases, when $E > {\cal O}(\sqrt{\lambda}/\rho)$ where $\lambda$ is the 't Hooft coupling of the ${\cal N}=4$ SYM.
This topology change affects the gluon scattering amplitude, and we find enhancement of the instanton amplitude as $\exp(\rho E)$ for the energy $E > {\cal O}(\sqrt{\lambda}/\rho)$.

The precise evaluation of the worldsheet area is generically difficult particularly when the worldsheet topology may change and also there exist multiple boundary conditions of different kinds, as in the present case. 
Although our model analysis is enough for our present purpose of finding a generic effect of the instantons to the amplitudes, the precise determination of the worldsheet configurations with different topology needs a sophisticated numerical method. In our companion paper \cite{ours2} we develop a machine learning method called physics-informed neural network (PINN) \cite{raissi2019physics} to be applied to generic minimal surfaces in curved geometry, and we use the method
to pin down the topology-changed configuration of the worldsheet studied in the present paper.

This paper is organized as follows. After we briefly review the Alday-Maldacena method \cite{Alday:2007he, Alday:2007hr} in Sec.~\ref{sec:2}, we
provide the picture of the topology change and the holographic order estimation of the instanton amplitude in Sec.~\ref{sec:3}. In Sec.~\ref{sec:4}, we provide a model analysis of the worldsheet to study the topology change and to discuss the existence of a minimal surface, in two models. The final Sec.~\ref{sec:sum} is for a summary and discussions. In App.~\ref{app:1}, we describe surfaces in global coordinates. In App.~\ref{app:2}, we discuss our worldsheet model limitation.


\section{A brief review of Alday-Maldacena method}
\label{sec:2}

First, let us briefly summarize the idea of Alday and Maldacena \cite{Alday:2007hr}.\footnote{See \cite{Alday:2008yw} for a detailed review.} The gluon scattering amplitudes correspond, in the holographically dual AdS gravity side, to an elongated string worldsheet since the gluon corresponds to a tiny string at the asymptotic past and the future. Then Alday and Maldacena took a T-duality along the $x^0, x^1, x^2, x^3$ directions to obtain a T-dual AdS geometry in which the boundary manifold is spanned not by the coordinates $(x^0, x^1, x^2, x^3)$ but by the momenta $(k_0, k_1, k_2, k_3)$ which are T-dual to the coordinates.
Eventually the T-dual geometry is again AdS${}_5$, whose radial coordinate is the inverse of the original radial coordinate.\footnote{This T-duality can be upgraded to fermionic T-duality \cite{Berkovits:2008ic} which is helpful to explore dual conformal symmetry of the SYM.} The original worldsheet is mapped again to another worldsheet in the T-dual AdS as the T-duality is a Lagrangian symmetry of the worldsheet theory. This worldsheet ends on the boundary of the T-dual AdS with a collection of null vectors ${\vec{k}}^{(i)}$
which are nothing but the momenta of the incoming and outgoing gluons, with the momentum conservation law $\sum_i \vec{k}^{(i)}=0$ meaning that the boundary of the worldsheet is a null polygon. The gluon scattering amplitude is proportional to $e^{-{\cal A}}$ where ${\cal A}$ is the area of the worldsheet. 

So, in summary, the $n$-gluon scattering amplitude with gluon momenta $\vec{k}^{(i)}$ ($i=1,2,\cdots, n$) in ${\cal N}=4$ SYM at large $N$ and strong coupling limit is proportional to $e^{-{\cal A}}$ where ${\cal A}$ is the area of the string worldsheet in the T-dual AdS${}_5$ spacetime which ends on polygon edges which are null vectors $\{\vec{k}^{(i)}\}$ at the boundary of the T-dual AdS${}_5$.

For later purposes, we summarize the T-duality procedure here.
The original AdS${}_5$ spacetime metric is 
\begin{align}
    ds^2 = \frac{R^2}{z^2}((dx^\mu)^2 + dz^2)
\end{align}
where $R$ is the AdS radius, and $z$ is the AdS radial coordinate ($z=0$ is the boundary of the AdS).
Now the T-duality along the directions $x^\mu$ $(\mu = 0,1,2,3)$ is taken, with the T-duality rule to get the T-dual coordinate $y^\mu$,
\begin{align}
    dy = \frac{R^2}{z^2} dx.
\end{align}
Then we arrive at 
\begin{align}
    ds^2 = \frac{R^2}{r^2}((dy^\mu)^2 + dr^2)
    \label{eq:T-dualAdS}
\end{align}
where the new radial coordinate $r \equiv R^2/z$ is defined.
Now we see that the T-dual spacetime is again an AdS${}_5$ spacetime.\footnote{This T-dual AdS geometry is a solution of the type IIB supergravity in 10 dimensions, not with the 5-form flux (as usually used for the AdS/CFT correspondence but with the nontrivial dilaton and the RR 0-form flux \cite{Kallosh:1998ji}.} Note that the radial coordinate is inverted: the original boundary is mapped to $r=\infty$ in the T-dual AdS, meaning that the boundary of the T-dual AdS $r=0$ corresponds to the IR limit of the CFT. In addition, the variable $y^\mu$ represents the momentum $k$, through the relation $y^\mu = 2 \pi \alpha' k^\mu$.

Let us just quote the final result of Alday and Maldacena \cite{Alday:2007he}. The gluon scattering amplitude is:
\begin{align}
    {\cal A}= \exp\left[iS_{\rm div} + \sqrt{\lambda}
    \frac{1}{8\pi}\left[\left(\log\frac{s}{t}\right)^2
    + c\right]
    \right]
\end{align}
where we have defined the Mandelstam variables
$s=-(k^{(1)} + k^{(2)})^2$ and $t = -(k^{(1)} + k^{(4)})^2$ and the constant number 
$c \equiv \left(\pi^2/3+2\log 2 - (\log 2)^2\right)/(4\pi)$, and 
$S_{\rm div}$ is a divergent contribution from the boundary part of the worldsheet.
Note that when $s=t$, the finite part of the amplitude does not depend on the energy of the gluons. It is just a constant proportional to $\sqrt{\lambda}$. The reason is the conformal invariance: there is no way to make this tunneling rate (the exponent, which is dimensionless) depend on the typical energy $E \sim \sqrt{-s}=\sqrt{-t}$ since there is no scale to cancel the $E$-dependence. The 'tHooft coupling dependence $\sqrt{\lambda}=R^2/\alpha'$ (according to the AdS/CFT dictionary)
is natural since the exponent is the Nambu-Goto action of the string worldsheet
\begin{align}
    S_{\rm NG} = \frac{1}{2\pi\alpha'} \int d^2\sigma \sqrt{-\det G}
\end{align}
where $G$ is the induced metric on the worldsheet, and as \eqref{eq:T-dualAdS} shows the induced metric is proportional to $R^2$, thus whatever the worldsheet shape depends on $E$, the resultant Nambu-Goto action should be proportional just to $R^2/\alpha'$ with no $E$-dependence.\footnote{In the context of Wilson loop and its dual in AdS \cite{Maldacena:1998im}, this fact is consistent with the standard result for the circular Wilson loop \cite{Berenstein:1998ij}: The expectation value does not depend on the radius of the circular Wilson loop, due to the conformal invariance.}

\section{Dual picture of the gluon scattering amplitude with instantons}
\label{sec:3}

\subsection{Instanton D3-brane and the worldsheet topology change}

\begin{figure}[t]
\centering
    \subfigure[Small instanton.]{\includegraphics[height=6.5cm]{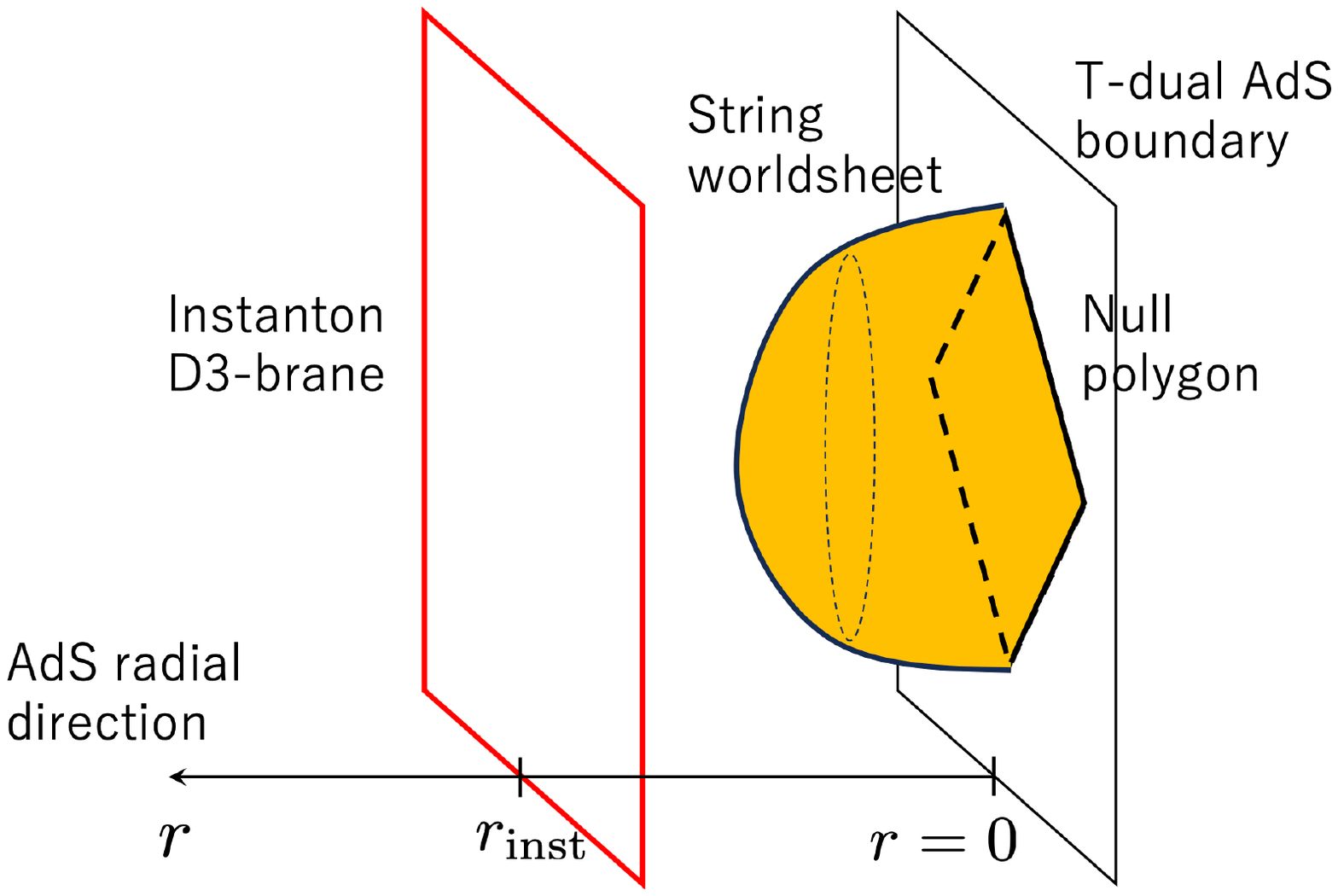}}
    \subfigure[Large instanton.]{\includegraphics[height=6.5cm]{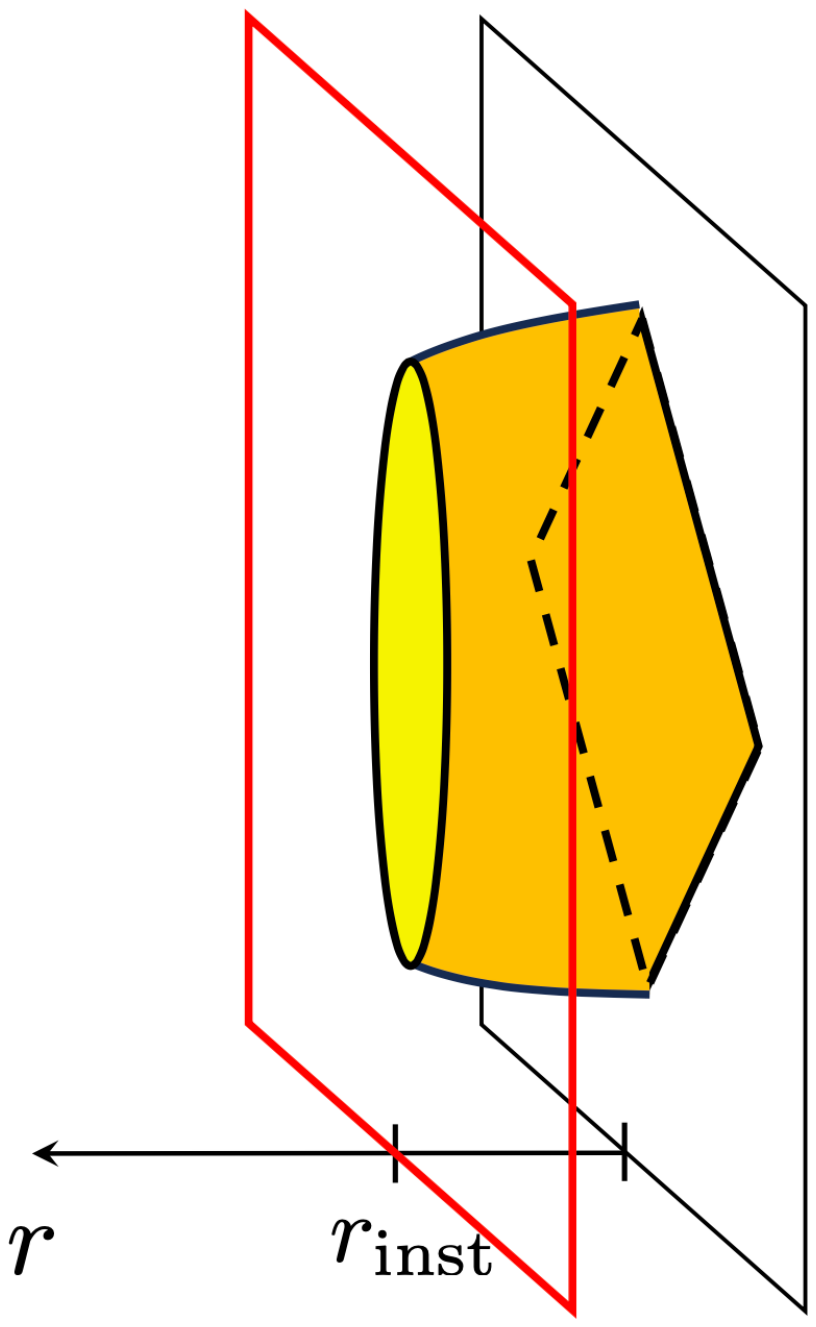}}
    \caption{The worldsheet configurations. Left: the instanton D3-brane is far away from the boundary of the T-dual AdS, corresponding to the small instanton. The string worldsheet is not affected by the D3-brane. Right: the instanton D3-brane is close to the boundary (the instanton is large), then the minimal surface has ends on the D3 so that it has a hole on the worldsheet.}
    \label{fig:topologychange}
\end{figure}

Now we are ready to study the effect of the instantons.\footnote{Once the Yang-Mills instanton is introduced to the gluon scattering amplitudes, the fermion zeromodes are necessary to have non-vanishing results. Here in this paper, we assume that a necessary number of those soft fermions (gauginos) are also included in the initial/final states of the scattering amplitudes.}
The Yang-Mills instanton in the ${\cal N}=4$ SYM corresponds to a
D-instanton in AdS \cite{Banks:1998nr}, according to the AdS/CFT dictionary. The evidence of this particular correspondence is seen as the fact that the moduli space metric of this SYM instanton is just the AdS${}_5$ geometry \cite{Dorey:1999pd, Bianchi:1998nk, Dorey:1998qh}. In the dictionary, it was found that the instanton size modulus $\rho$ corresponds to the radial coordinate of the location of the D-instanton: $z_{\rm inst} = \rho$. Thus in the T-dual AdS coordinate, we find
\begin{align}
    r_{\rm inst} = \frac{R^2}{\rho}.
\end{align}

After taking the T-duality along the $x^\mu$ directions, the D-instanton is mapped to a D3-brane which is parallel to the boundary of the T-dual AdS${}_5$. We call this an ``instanton D3-brane."
This D3-brane has an important effect on the worldsheet of the string for the gluon scattering amplitudes: the worldsheet can now end on the instanton D3-brane.
Whether the worldsheet actually ends on the instanton D3-brane or not actually depends on the location of the instanton D3-brane. When the D3 is close to the boundary of the T-dual AdS, then the worldsheet which is the minimal surface prefers to have ends on the D3, while when the D3 is far away from the boundary of the T-dual AdS, the worldsheet is never affected by the D3. Therefore, when we move the location of the instanton D3-brane, there occurs a topology change of the string worldsheet. See Fig.~\ref{fig:topologychange}.\footnote{This topology change is similar to what happens in the case of quark scattering in the holographic dual of the ${\cal N}=2$ supersymmetric QCD \cite{Komargodski:2007er,McGreevy:2007kt}. The introduced flavor D7-brane is T-dualized to a Euclidean D3-brane, and the worldsheet minimal surface can end on the D3-brane. The orientation of our instanton D3-brane is different from that.}

Note that this kind of topology change of the minimal surfaces in AdS/CFT is a popular situation. Even without a precise determination of the worldsheet configuration, we can infer easily through the other popular topology change in the AdS/CFT examples: Wilson loops and Debye screening, chiral symmetry breaking, entanglement entropy transition, and unitarity restoration / Page curve --- all of these examples correspond respectively to a particular kind of minimal surface and its topology change --- the string worldsheet at a black hole horizon, the flavor D-brane at the horizon, the Ryu-Takayanagi surface, and the island near the horizon; see Table \ref{tab:AdS_CFT_dictionary}. 

\begin{table}[t]
\centering
\begin{tabular}{c|c|c|c}
\begin{tabular}{c}
Worldvolume \\ dim.
\end{tabular}
& Surface in AdS$_{d+1}$ & CFT$_d$ quantity 
&
\begin{tabular}{c}
Surface topology change \\ means in CFT$_d$
\end{tabular}
\\
\hline\hline
 $1+1$ & String worldsheet & Wilson loop \cite{Maldacena:1998im} & Debye screening \cite{Rey:1998bq}\\
\hline
 $(d-1)+1$ & D-brane & Quark sector \cite{Karch:2002sh} & Symmetry breaking \cite{kruczenski2004towards}\\
\hline
 $(d-1)+0$ & 
\begin{tabular}{c}
 Ryu–Takayanagi \\
 surface
\end{tabular}
& Entanglement \cite{Ryu:2006bv} & 
\begin{tabular}{c}
Entanglement \\
transition 
\cite{klebanov2008entanglement} 
\end{tabular}
\\
\hline
 $d+0$ & Volume & 
\begin{tabular}{c}
  Computational \\
complexity \cite{Susskind:2014rva} 
\end{tabular}
& Page curve  \cite{Penington:2019npb,Almheiri:2019psf} \\
\end{tabular}
\caption{A popular dictionary between extremal surfaces in AdS$_{d+1}$ geometries and CFT$_d$ observables, and the CFT meaning of the topology change of the surface.}
\label{tab:AdS_CFT_dictionary}
\end{table}

For example, let us look at 
the situation of the Wilson loops at finite temperature. The AdS dual of it is a minimal surface worldsheet in the presence of the black hole horizon, and depending on the location of the horizon, the worldsheet configuration experiences the topology change \cite{Rey:1998bq}. In the dual SYM side, this phenomenon corresponds to the Debye screening. This ``phase transition" is the first order.\footnote{
Once a quantum fluctuation of the worldsheet is taken into account, the phase transition may become a higher order. There is a discussion \cite{Faulkner:2008hm} that this kind of string/brane touching the horizon could be the third-order transition. It is amusing to note that the Machine learning stochastic gradient descent which we use to find the minimal surface looks like the quantum fluctuation to find a better minimum.
} The other examples listed in Table \ref{tab:AdS_CFT_dictionary}, and many other similar examples in AdS/CFT, are well-understood.

Minimal surfaces, or extremal surfaces in general, are important everywhere in string theory and in general relativity, simply because, due to the general coordinate transformation symmetry, the only viable physical specification of observables is extended objects anchored at some boundary of the gravitational spacetime. The importance is vividly seen in examples of the AdS/CFT correspondence \cite{Maldacena:1997re} as the volume (area) of the minimal surfaces corresponds to important quantities in the dual quantum field theories, as listed in Table \ref{tab:AdS_CFT_dictionary}.
Depending on the dimensionalities and how the extremal surfaces are arranged in the curved spacetimes, the corresponding physical observables in the CFT side range in variety. Therefore, the topology change of minimal surfaces in curved geometry shows up generally in AdS/CFT, and our finding in this paper adds one novel example of the gluon scattering amplitudes.

\subsection{Instanton amplitude}

Let us evaluate the effect of the instanton D3-brane on the gluon scattering amplitude.
When the worldsheet ends on the D3-brane, there is a hole on the worldsheet. The size of the hole is roughly of the order of $2\pi\alpha' E$ where $E$ is the typical order of the gluon momenta, since this is the size of the boundary null polygon. The proper length of the circumference of the hole is $\sim \alpha' E \sqrt{G_{yy}}$. Then we naively expect that the change of the Nambu-Goto action under the slight change of the D3-brane radial location is
\begin{align}
    \delta S_{\rm NG}
    \sim \alpha' E \sqrt{G_{yy}} \cdot \sqrt{G_{rr}} \delta r_{\rm inst}.
\end{align}
Using the metric \eqref{eq:T-dualAdS} and making an integration of this equation over $r_{\rm inst}$, we find \footnote{This simple model is studied in Sec.~\ref{subsec:model1} in detail.}
\begin{align}
    S_{\rm NG} \sim c_1 \sqrt{\lambda} - c_2 E \frac{R^2}{r_{\rm inst}}
    = c_1 \sqrt{\lambda}- c_2 \rho E,\label{SNG_naive}
\end{align}
where $c_1$ and $c_2 (>0)$ are ${\cal O}(1)$ quantities. The constant part $c_1$ should include a divergent part, but our important finding is the $\rho$ and $E$ dependence which appears in the second term, which has no $\lambda$ dependence.

\begin{figure}[t]
\centering
   \includegraphics[height=8cm]{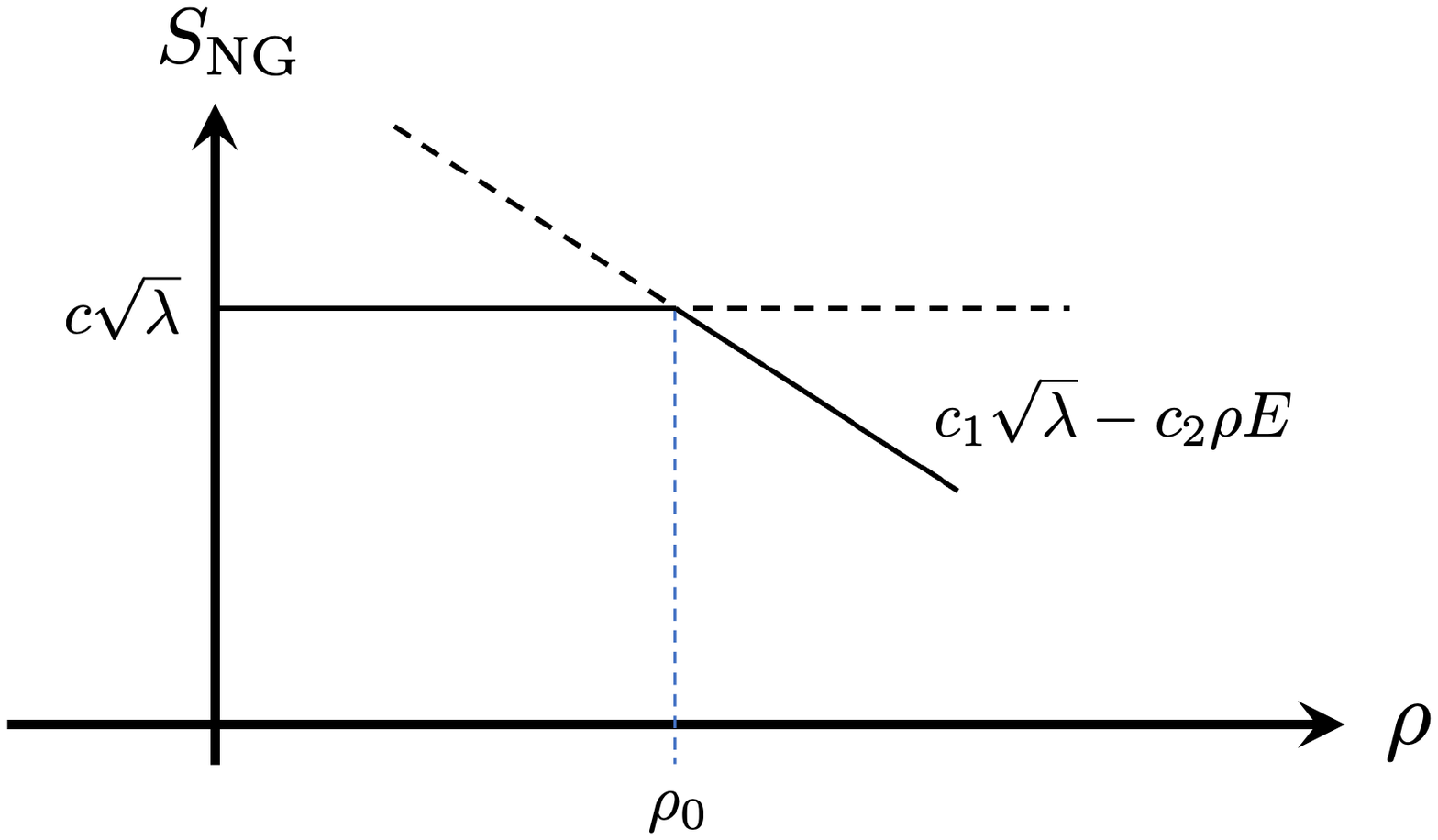}
    \caption{The worldsheet area $S_{\rm NG}$ as a function of the instanton size modulus $\rho$. It goes through a topology change.}
    \label{fig:SNG_org}
\end{figure}

Putting all findings together, we find that there exists a phase transition size of the instanton $\rho_0$ for a fixed gluon energy $E$, and the instanton amplitude behaves as
\begin{align}
    & {\cal A} \sim \exp\left[-\frac{8\pi^2}{g_{\rm YM}^2}\right]
    \exp\left[- c \sqrt{\lambda}\right] & (\rho<\rho_0) \label{eq:A1}\\    
    & {\cal A} \sim \exp\left[-\frac{8\pi^2}{g_{\rm YM}^2}\right]
    \exp\left[- c_1 \sqrt{\lambda}+ c_2 \rho E\right] & (\rho>\rho_0) \label{eq:A2}
\end{align}
We have included the D-instanton tension $\frac{2\pi}{g_s} = \frac{8\pi^2}{g_{\rm YM}^2}$ where $g_s$ is the string coupling related to the Yang-Mills coupling constant as $g_{\rm YM}^2 = 4\pi g_s$. And, since the topology change should occur, we need $c_1 > c$. 
See Fig.~\ref{fig:SNG_org} for the behavior of the exponent $S_{\rm NG}$ which is related to the amplitude as ${\cal A} \sim \exp[-S_{\rm NG}]$. 

Due to the balance between the two expressions, we find the critical size of the instanton for the phase transition due to the worldsheet topology change for fixed $E$ as
\begin{align}
    \rho_0 =c_3 \frac{\sqrt{\lambda}}{E} 
\end{align}
where $c_3$ is an ${\cal O}(1)$ positive number.

Let us look at the results \eqref{eq:A1} and \eqref{eq:A2} in a different manner; for a fixed instanton modulus $\rho$, if we vary the energy, then the gluon scattering amplitudes behave as
\begin{align}
    & {\cal A} \sim \exp\left[-\frac{8\pi^2}{g_{\rm YM}^2}\right]
    \exp\left[- c \sqrt{\lambda}\right] & (E<E_0) \label{eq:A1'}\\    
    & {\cal A} \sim \exp\left[-\frac{8\pi^2}{g_{\rm YM}^2}\right]
    \exp\left[- c_1 \sqrt{\lambda}+ c_2 \rho E\right] & (E>E_0) \label{eq:A2'}
\end{align}
That is to say, the exponential growth \eqref{eq:A2} of the amplitude in energy $E$ for a fixed size $\rho$ of the instanton occurs for  
\begin{align}
E> E_0(\rho) \equiv c_3 \sqrt{\lambda}/\rho  
\label{E>rho0}
\end{align}
where $E_0$ is the critical energy at which the phase transition occurs and beyond which the exponential growth of the amplitude starts.
In summary, we find in \eqref{eq:A1'} and \eqref{eq:A2'} that, for a fixed instanton size modulus $\rho$, the instanton amplitude for high $E$ is exponentially enhanced. This growth is only for the high $E$ region, and there appears a phase transition at a certain gluon energy \eqref{E>rho0}.

In the next section, we provide an estimation of the worldsheet configuration and its area in the presence of the instanton D3-brane, and will qualitatively confirm the phase transition structure described above. As we described in the introduction, the worldsheet configuration of Fig.~\ref{fig:topologychange}(b) contains a combination of different boundary conditions, and the numerical evaluation is difficult. Thus we resort to several ansatz to look at the topology change.\footnote{As we shall discuss in the final section, the numerical evaluation of the topology-changed minimal surface has a number of difficulties. In our companion paper \cite{ours2}, we introduce an AI-assisted method called physics-informed neural network (PINN) \cite{raissi2019physics} to discover the configuration of the topology-changed minimal surfaces. It provides further evidence of the topology change.}

\subsection{Multiple instantons}

In this subsection, we study the amplitudes in the presence of multiple instantons. Obviously, as prescribed in earlier subsections, if we take the T-duality, all the D-instantons are mapped to parallel instanton D3-branes in the T-dual AdS${}_5$ geometry. The original moduli parameters of the Yang-Mills 't Hooft instantons are the set $\{\rho_i, x^\mu_i\}$ which is the size and the location of each instanton labeled by $i=1, \cdots, n$ where $n$ is the instanton number. After the T-duality, we have the parallel D3-branes whose radial location is specified by 
\begin{align}
    \left\{\frac{R^2}{\rho_1}, \frac{R^2}{\rho_1}, \cdots, \frac{R^2}{\rho_n} 
    \right\} .
    \label{eq:colr}
\end{align}
Note that the information $x^\mu_i$ is gone due to the T-duality.

\begin{figure}[t]
\centering
   \includegraphics[height=6cm]{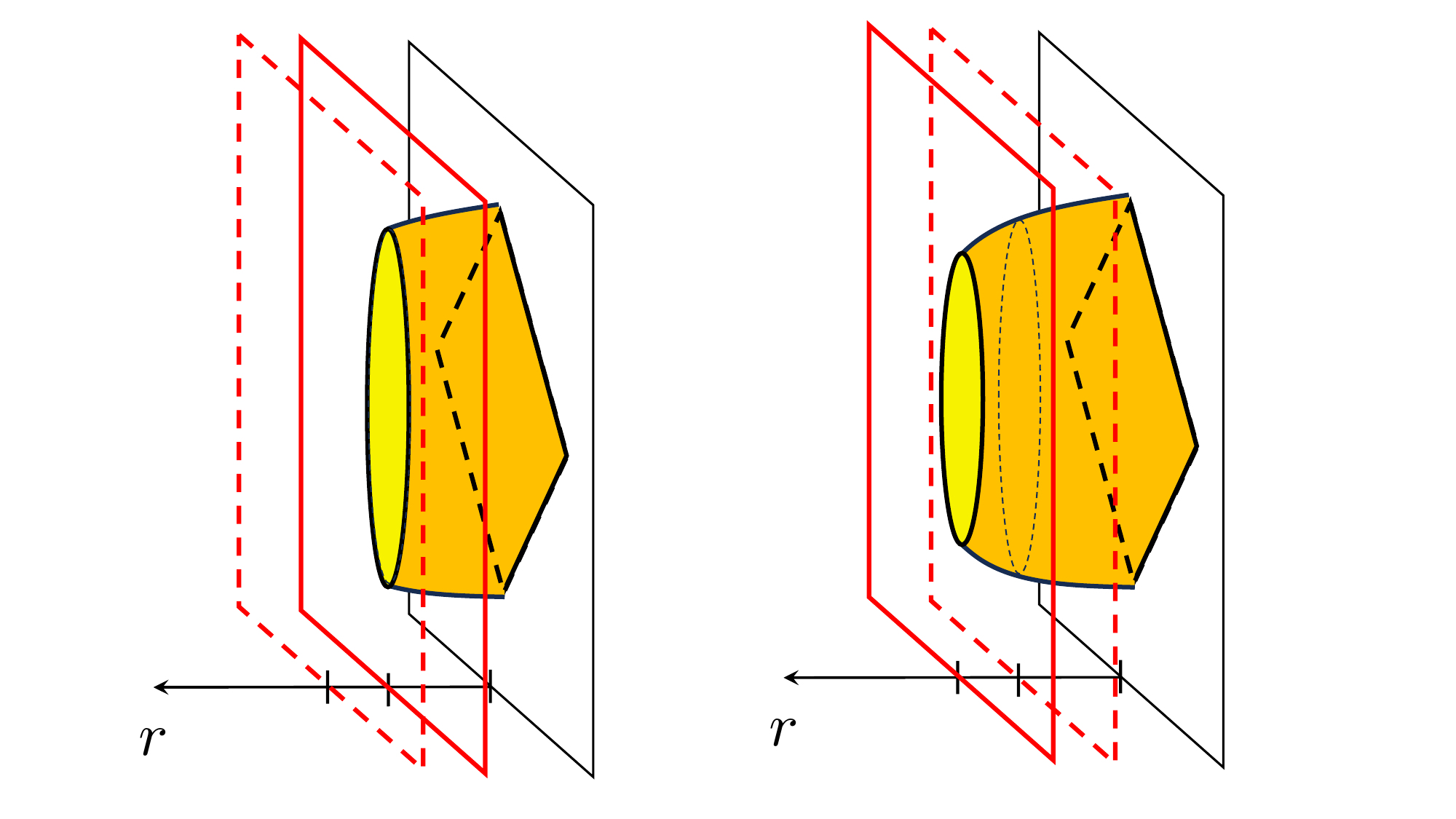}
    \caption{The case of two instantons. (Left) The largest instanton contributes the instanton amplitude. The solid plane is the instanton D3-brane with the largest instanton size. (Right) Other smaller instantons also contribute as less-important saddle points.}
    \label{fig:Multi}
\end{figure}

When we put the string worldsheet, there is a choice of which D3-brane the worldsheet ends on. Since the amplitude is bigger for a smaller area of the worldsheet, the dominant contribution should come from the worldsheet which ends on the D3-brane closest to the t-dual AdS boundary. See Fig.~\ref{fig:Multi} (Left). This means that the smallest value among \eqref{eq:colr} is selected as the dominant contribution, thus the largest instanton is automatically chosen.

We conclude that the gluon scattering amplitude for a fixed set of instantons is the same as \eqref{eq:A1'} and \eqref{eq:A2'}, with the phase transition energy defined by
\begin{align}
E> E_0 \equiv c_3 \sqrt{\lambda}/\rho_{\rm largest}  
\label{E>rho0Multiple}
\end{align}

Of course, a smaller contribution coming from other D3-branes exists (see Fig.~\ref{fig:Multi} (Right)) because the worldsheet can penetrate D-branes. However, their contributions are suppressed by a factor $\exp[c_2 (\rho-\rho_{\rm largest})E]$.

In summary, we find that even in the case of multiple instantons, only the largest instanton contributes to the gluon scattering amplitude, and the expression of the amplitude reduces to that of the case with a single instanton.


\section{Model estimation of the minimal surface and topology change}
\label{sec:4}

The aim of this section is to provide a model-based estimation that supports the naive estimation of the minimal surface $S_{\rm NG}$ in the presence of an instanton D3-brane \eqref{SNG_naive}.

\subsection{Worldsheet parameterization}

The starting point is to define the boundary conditions for the worldsheet. Since we perform a T-dual transformation on the usual AdS space, the boundary condition is now specified not by real-space coordinates $(x_0,x_1,x_2,x_3)$ but by momentum-space coordinates $(y_0,y_1,y_2,y_3)$. In the situation of 2-to-2 gluon scattering, the worldsheet has to end on the null polygon on the T-dual AdS boundary, which is composed of momentum vectors corresponding to each in- and out-state of two gluons. As in \cite{Alday:2007hr}, we set these momenta as follows:
\begin{align*}
    &k_1=2\left(\begin{array}{r}1\\0\\1\\0\end{array}\right)\ ,
    &&k_2=2\left(\begin{array}{r}-1\\-1\\0\\0\end{array}\right)\ ,
    &&k_3=2\left(\begin{array}{r}-1\\0\\-1\\0\end{array}\right)\ ,
    &&k_4=2\left(\begin{array}{r}1\\1\\0\\0\end{array}\right) \ ,
\end{align*}
with AdS radius being set to $R=1$. These vectors sum to zero, satisfying momentum conservation. From this point on, we omit the  $y_3$ component, as it is always zero. The ordering of these momenta is crucial because it corresponds to the color ordering in the SYM amplitude. In the absence of an instanton D3-brane, any other choice of momenta can be related by conformal transformations. Even with the instanton D3-brane present, the quantity we wish to estimate depends only on the ratio of the gluon energy scale $|\vb{k}|$ and the inverse instanton size $1/\rho$, so we simply fix these momentum vectors and use $\rho=R^2/r_{\rm inst}$ as an order parameter. 

The polygon formed by these momentum vectors is embedded on the AdS boundary ( at $r=0$) in such a way that the four vertices are placed at:
\begin{align}
    (y_0,y_1,y_2)=(1,1,1)\ , (-1,-1,1)\ , (1,-1,-1)\ , (1,1,-1)\ .
\end{align}
This shape appears crown-like in the $(y_0,y_1,y_2)$ space. While in the $(y_1,y_2,r)$ space, the edges of the polygon form a $2\times2$ square on the $y_1$-$y_2$ plane, and the worldsheet will extend into AdS space $r>0$ to close itself or reach the instanton D3-brane. 

Having established these boundary conditions, we proceed to solve the equation of motion for the worldsheet. We use $y_1$ and $y_2$ as worldsheet coordinates, as a choice of the gauge.
This parametrization allows the worldsheet configuration to be described by two real functions, $y_0(y_1,y_2)$ and $r(y_1,y_2)$. The action we must minimize is:
\begin{align*}
    S=\iint dy_1 dy_2\frac{1}{r^2}\sqrt{1+\left(\partial_1r\right)^2+\left(\partial_2r\right)^2-\left(\partial_1y_0\right)^2-\left(\partial_2y_0\right)^2 -\left(\partial_1r\partial_2y_0-\partial_2r\partial_1y_0\right)^2}
\end{align*}
where $\partial_1$ and $\partial_2$ represent $\pdv{y_1}$ and $\pdv{y_2}$. In the absence of the instanton, or the D3-brane in the T-dual AdS spacetime, an analytical solution of this problem was found by Alday and Maldacena \cite{Alday:2007hr}:
\begin{align}
    &r(y_1,y_2)=\sqrt{(1-y_1^2)(1-y_2^2)}\ ,  && y_0(y_1,y_2)=y_1y_2\ , \ \
    \label{eq:1cusp}
\end{align}
hereinafter referred to as $r^{\rm AM}$ and $y_0^{\rm AM}$. This solution can be obtained by rotating another minimal surface terminating on the boundary null cusp $y_0=\pm y_1$ \cite{Ooguri1999}: 
\begin{align}
    &y_0(\tau,\sigma)=e^{\tau}\cosh\sigma \ ,
    &&y_1(\tau,\sigma)=e^{\tau}\sinh\sigma\  ,
    && r(\tau,\sigma)=e^{\tau}\sqrt{2} \ ,
    && y_2=y_3=0 \label{eq:1cusp_ts}
\end{align}
using AdS$_5$ isometries (see Appendix \ref{app:1}). This surface resembles an elliptic hemi-cone, 
extending to the infinity of Poincar\'e patch $r\rightarrow \infty$. The other three cusps of the polygon are hidden away to this infinity. Consequently, one can appeal to scale invariance, which corresponds to a constant $\tau$ shift in \eqref{eq:1cusp_ts}, and can reduce the search for the minimal surface to a one-dimensional problem.

In our scenario, however, D3-brane at $r=r_{\rm inst}$ breaks the scale invariance, so the strategy mentioned above is not applicable. Furthermore, the worldsheet can now terminate on the D3-brane and change its topology, so we must incorporate additional boundary conditions: a Dirichlet condition on $r=r_{\rm inst}$, and a Neumann condition for $y_0,y_1,y_2,y_3$. Moreover, the shape of the endline of the worldsheet on the D3-brane cannot be decided in advance, so we also have to optimize the location where these conditions are imposed. Such a system, with its moving boundary conditions, is analytically difficult to solve, recalling examples such as the soap film problem. For these reasons, we decided to use variational methods to approximate the minimal surface and calculate its area. 

In the subsequent subsections, we introduce the ansatz of the worldsheet which we optimize, and present the Nambu-Goto area of the approximated minimal surface. Based on the result, we extract the dependence of the area on the instanton size $\rho_{\rm inst}=1/r_{\rm inst}$, and estimate the switching point where the configurations with a hole become more dominant than the configuration without a hole (see figure 1).

Before we get into the main topic, there is one thing to note. The Nambu-Goto area of a surface extending to the AdS boundary is inherently divergent, thus requiring some regularization. In \cite{Alday:2007hr}, which considers the same system but without the D-brane, the Nambu-Goto area is regularized by using dimensional regularization along the coordinate $r$. However, in our case, to observe the topology change from a surface without a hole to one with a hole, it suffices to estimate the difference in their areas, which is expected to be finite.

\subsection{Model 1: Simple cut of the original surface}
\label{subsec:model1}

As a zeroth-order approximation, we first numerically calculate the area of the minimal surface simply cut by the plane $r=r_{\rm inst}$. The model analysis provided in the previous section is a linear approximation of this cut model in this subsection.

\begin{figure}[t]
\centering
   \includegraphics[height=8cm]{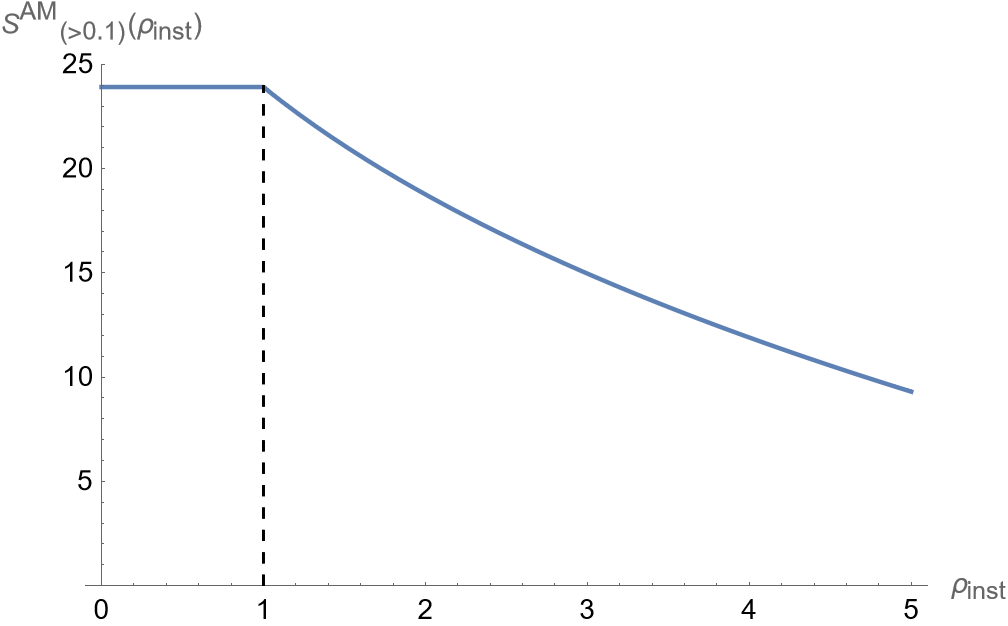}
    \caption{The worldsheet area of the no-hole minimum surface simply cut by $r=r_{\rm inst}=1/\rho_{\rm inst}$. The topology of the worldsheet changes at $r_{\rm inst}=1$ .}
    \label{fig:SNG}
\end{figure}

In this case, the worldsheet configuration is described by functions $r^{\rm AM}$ and $y_0^{\rm AM}$:
\begin{align}
    & Y=\mqty(y_0^{\rm AM}\\y_1\\y_2\\r^{\rm AM}) \ ,
    && y_0^{\rm AM}(y_1,y_2)=y_1y_2 \ ,
    && r^{\rm AM}(y_1,y_2)=\sqrt{(1-y_1^2)(1-y_2^2)}
\end{align}
but is restricted by the condition $r<r_{\rm inst}$. The background metric is the (T-dual) AdS metric, given by
\begin{align}
    ds^2=G_{\mu\nu}dY^\mu dY^\nu =\frac{-dy_0^2+dy_1^2+dy_2^2+dr^2}{r^2} \ .
\end{align}
Consequently, the resulting Euclidean action is:
\begin{align}
    S&=\iint_{r(y_1,y_2)<r_{\rm inst}} \sqrt{
      \left( G_{\mu\nu}\pdv{Y^\mu}{y_1}\pdv{Y^\nu}{y_1}\right)
      \left( G_{\rho\sigma}\pdv{Y^\rho}{y_2}\pdv{Y^\sigma}{y_2}\right)
      -\left( G_{\mu\nu}\pdv{Y^\mu}{y_1}\pdv{Y^\nu}{y_2}\right)^2
    } \nonumber \\
    &=\iint_{r(y_1,y_2)<r_{\rm inst}} \frac{dy_1 dy_2 }{(1-y_1^2)(1-y_2^2)} \ .
\end{align}
In the $(y_1,y_2,r)$ space, the worldsheet defined by $r=r^{\rm AM}$ forms a dome-like surface with an upper bound $r\leq1$. Therefore, setting $r_{\rm inst}>1$ leads to
\begin{align}
    S^{\rm AM}=\iint \frac{dy_1 dy_2 }{(1-y_1^2)(1-y_2^2)} \ ,
\end{align}
an integral previously considered in the literature \cite{Alday:2007hr}. Since the bare values of both $S$ and $S^{\rm AM}$ are divergent, we introduce a cutoff by considering only the contribution from the region $r>\tau_0$, which gives us the regularized action
\begin{align}
    S_{>\tau_0}=\iint_{\tau_0<r<r_{\rm inst}} \frac{dy_1 dy_2 }{(1-y_1^2)(1-y_2^2)} \ .
\end{align}  
By applying the same cutoff procedure to the surface area without a hole $S^{\rm AM}$, the difference  $S_{>\tau_0}-S^{\rm AM}_{>\tau_0}$ becomes independent of the choice of the cutoff parameter $\tau_0$. Thus, we set $\tau_0=0.1$ from now on. Fig.~\ref{fig:SNG} shows the result of numerical integration of $S_{>\tau_0}$ for each $r_{\rm inst}$\footnote{These integrals mentioned above cannot be written in terms of elementary functions.}. We see that the behavior of the worldsheet area $S$ as a function of $r_{\rm inst}$ for fixed gluon momentum, shown in Fig.~\ref{fig:SNG}, is qualitatively coincident with that in the previous section, Fig.~\ref{fig:SNG_org}.

\subsection{Model 2: Surface partially satisfying the D-brane boundary condition }
\label{subsec:model2}

Since with the simple cut the worldsheet configuration doesn't satisfy the boundary condition on the instanton D3-brane, we better modify the original functions $r^{\rm AM}$ and $y_0^{\rm AM}$ to create a more proper ansatz. We expect that the resulting plot of the Nambu-Goto area $S$ will be shifted downwards and that the phase transition point will also be shifted to the region $r_{\rm inst}>1$. This is because even if the location of the brane $r=r_{\rm inst}$ is higher than the unity, it is possible for the worldsheet to cost less area by reaching the D3-brane to terminate than by closing itself to be the no-hole solution $r=r^{\rm AM}$.  

In principle, a complete solution would require modifying both $r^{\rm AM}$ and $y_0^{\rm AM}$. However, due to the analytical complexity involved, we adopt a simplified approach in the next subsection: we modify only the T-dual AdS depth $r$ to satisfy the Dirichlet boundary condition while keeping the function $y_0^{\rm AM}$ unchanged from its original form.\footnote{The proper treatment of the Neumann boundary condition for all the coordinates is studied with the use of the AI method PINN, in our companion paper \cite{ours2}.}


Let us construct a configuration that satisfies the boundary condition on $r$ and then optimize its shape to minimize the Nambu-Goto area. First and foremost, the minimal surface with a hole on the D3-brane must adhere to the symmetries of the boundary condition:
\begin{itemize}
    \item $y_0$ is an odd function of both $y_1$ and $y_2$\ .
    \item $r$ is an even function of both $y_1$ and $y_2$\ .
    \item Both $y_0$ and $r$ are invariant under exchanging $y_1$ and $y_2$\ .
\end{itemize}
Based on these prerequisites, for simplicity, we impose the following restriction on the configuration:
\begin{itemize}
    \item $y_0$ remains to be $y^{\rm AM}=y_1y_2 $\ .
    \item $r$ is a function only on $\tau\equiv \sqrt{(1-y_1^2)(1-y_2^2)}=r_{\rm AM}$, resulting in that the hole on the D3-brane is a contour line of $r^{\rm AM}$.
    \item $r$, as a function of $\tau$, approaches the AdS boundary in the same manner as the solution without a hole :
    \begin{align*}
        &r \rightarrow r^{\rm AM}=\tau\ , 
        && \text{as} \ \ \ \tau=\sqrt{(1-y_1^2)(1-y_2^2)}\rightarrow 0
    \end{align*}
\end{itemize}
\begin{figure}[t]
\centering
   \includegraphics[height=8cm]{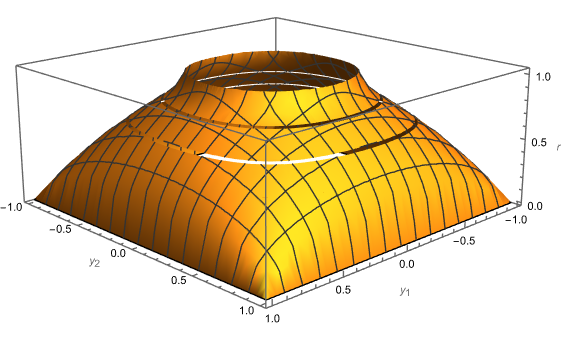}
    \caption{A worldsheet configuration of the ansatz \eqref{eq:rmodifdef} with the parameters $(\tau_1,\tau_2,\tau_3)=(0.65,0.80,0.88)$, reaching the instanton D3-brane placed at $r_{\rm inst}=1.1$. In this example, the regularized Nambu-Goto area (cut off by $\tau>\tau_0=0.1$) is 23.74, which is smaller than the area of the surface without a hole $S^{\rm AM}=23.91$ (with the same cutoff).}
\label{fig:r_vs_y1_y2_065_080_088_110}
\end{figure}
Respecting these requirements, we propose the following ansatz for the function $r$:
\begin{align}
     &0<\tau<\tau_1:\ \ && r(\tau) =\tau \nonumber \\
     &\tau_1<\tau<\tau_2:\ \ 
     && r(\tau) =\frac{\tau^2+\tau_1^2}{2\tau_1}
     \label{eq:rmodifdef}     
\\
     &\tau_2<\tau<\tau_3:\ \ 
&& r(\tau) =r_{\rm inst}-\sqrt{\tau_3^2-\tau^2}\sqrt{a(\tau_1,\tau_2,\tau_3)\tau^2+b(\tau_1,\tau_2,\tau_3)} \nonumber
\end{align}
The curve $r$ is a smooth patchwork of quadratic curves in the $(\tau^2,r)$ plane. To be more specific, $r$ coincides with $r^{\rm AM}$ in the region $0<\tau<\tau_1$, is linear in $\tau^2$ in the region $\tau_1<\tau<\tau_2$, and then is bent upward to be a quadratic curve in the region $\tau\in(\tau_2,\tau_3)$ to meet the D3-brane located at $r=r_{\rm inst}$. $a$ and $b$ are constants determined by requiring the curve $r$ and its first derivative to be continuous at $\tau=\tau_2$:
\begin{align*}
    a(\tau_1,\tau_2,\tau_3,r_{\rm inst})
    &=\frac{1}{(\tau_3^2-\tau_2^2)^2}\left(r_{\rm inst}-\frac{\tau_2^2+\tau_1^2}{2\tau_1}\right)
    \left(r_{\rm inst}-\frac{2\tau_3^2-\tau_2^2+\tau_1^2}{2\tau_1}\right) \ ,
    \\
    b(\tau_1,\tau_2,\tau_3,r_{\rm inst})
    &=\frac{1}{(\tau_3^2-\tau_2^2)^2}
    \left(r_{\rm inst}-\frac{\tau_2^2+\tau_1^2}{2\tau_1}\right)
    \left((\tau_3^2-2\tau_2^2)r_{\rm inst}+\frac{\tau_3^2\tau_2^2-\tau_3^2\tau_1^2+2\tau_2^2\tau_1^2}{2\tau_1}\right) \ .
\end{align*}
The sign of $a$ corresponds to whether the curve $r$ in $\tau_2<\tau<\tau_3$ is hyperbolic ($a>0$), parabolic ($a=0$) or elliptic ($a<0$). 


\begin{figure}[t]
\centering
   \includegraphics[height=8cm]{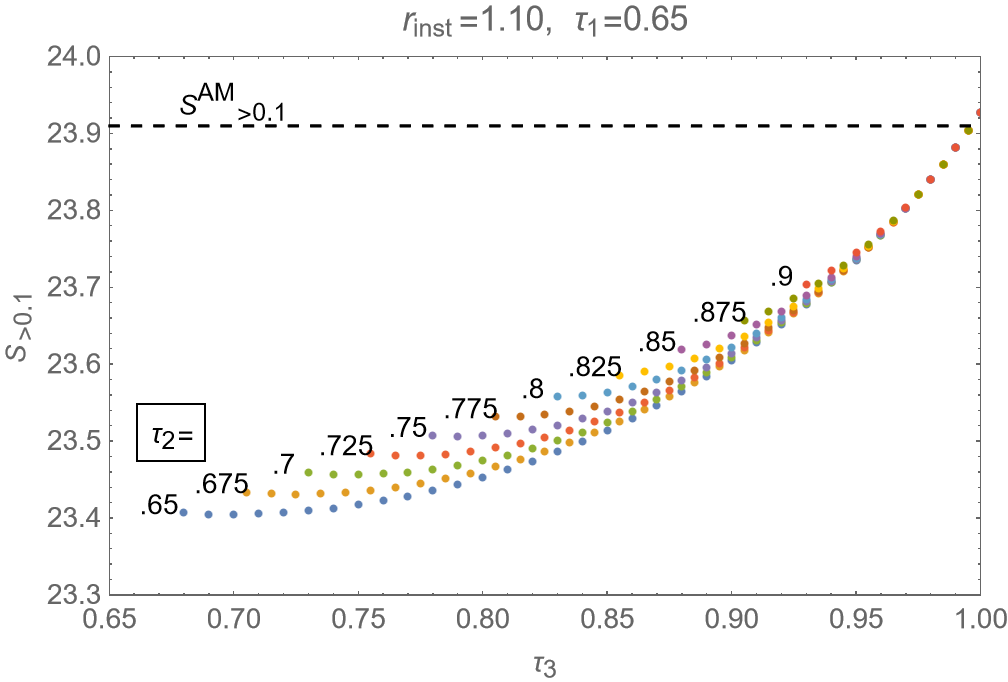}
    \caption{Regularized worldsheet area $S_{>0.1}$ of $r$ modified from $\tau>\tau_1=0.65$, reaching the instanton D3-brane placed at $r_{\rm inst}=1.1$. The horizontal dashed line shows the value of $S^{\rm AM}_{>0.1}$ which is given by the minimum surface without a hole.}
    \label{fig:SNG_r110_t065}
\end{figure}

To find non-trivial spacelike minimal surfaces, the ansatz \eqref{eq:rmodifdef} we introduce is designed to extend inward $(1-y_1^2)(1-y_2^2)<0$ from the AdS boundary, as depicted in Fig.~\ref{fig:r_vs_y1_y2_065_080_088_110}. This shape is quite natural, as expected from the standard minimal surfaces in flat spacetime, or from the behavior of Wilson loops in AdS/CFT. The justification of this model will be studied in our companion paper \cite{ours2} in which we will resort to an AI-assisted numerical method to find a solution that looks exactly like Fig.~\ref{fig:r_vs_y1_y2_065_080_088_110}.

As a note we shall comment on a trivial solution with zero Nambu-Goto area, consisting of null surfaces. In the $(y_1,y_2,r)$ space, this null surface appears as a surface of a square prism, extending straight from $r=0$ to $r=r_{\rm inst}$.  However, we discard such null solutions and instead seek minimal surfaces with non-zero area, due to the following reasons: First, the null solutions consist of null surfaces but at the intersection of the null surfaces the equations of motion are not satisfied. Second, as the null solution would correspond to a non-interacting process where two gluons pass through each other. In our method, this situation is reflected by the fact that null surfaces contribute nothing to the amplitude $\mathcal A $, which is estimated as $\mathcal A \sim e^{-S}$. 

Within the model ansatz \eqref{eq:rmodifdef}, the model parameter space spanned by $(\tau_1, \tau_2, \tau_3)$ needs to be scanned. More precisely, 
for the fixed instanton size $\rho_{\rm inst}$, or the D3-brane distance $r_{\rm inst}$, the parameters $(\tau_1,\tau_2,\tau_3)$ are varied within the range $\tau_0<\tau_1<\tau_2<\tau_3<1$, and the regularized Nambu-Goto area $S_{>\tau_0}$ is calculated to find its minimum value. 
Because scanning the three-dimensional parameter space numerically is quite costly, here we present a search result of a one-parameter slice in the model space. Several critical issues on the total space scan are discussed in detail in Appendix \ref{app:2}.

Fig.~\ref{fig:SNG_r110_t065} shows the result with fixed parameters $\tau_1=0.65$ and $r_{\rm inst}=1.10$. In this case, for each fixed $\tau_2$, there exists a minimum point $\tau_3$ located slightly over or at the $\tau_2$. 
We find that the minimum value of the Nambu-Goto action is lower than that of Alday-Maldacena, suggesting that the configuration with the hole through the topology change is favored to provide a bigger scattering amplitude.


\section{Summary and discussions}
\label{sec:sum}

In this paper, we introduce the effect of Yang-Mills instantons in the evaluation of the gluon scattering amplitudes in ${\cal N}=4$ SU($N$) SYM at large 't Hooft coupling $\lambda$ and large $N$, by generalizing the holographic calculation by Alday and Maldacena. The instanton corresponds to a D3-brane in the T-dual AdS${}_5$ spacetime in which a string worldsheet ending on a null polygon at the AdS boundary is placed. Due to the instanton D3-brane, the string worldsheet may undergo a topology change --- the string can end on the instanton D3-brane. This is reminiscent of the direct interaction between the gluon and the Yang-Mills instanton, as expected. 

The topology change drastically modifies the scattering amplitude, and we find that the energy dependence of the instanton amplitude \eqref{eq:A2} is $\exp[c_2 \rho E]$ where $\rho$ is the size of the Yang-Mills instanton, when $E> c_3 \sqrt{\lambda}/\rho$ \eqref{E>rho0}. On the other hand, when $E< c_3 \sqrt{\lambda}/\rho$, the instanton amplitude has no energy dependence \eqref{eq:A1}, and reduces to that of Alday-Maldacena except for the instanton factor $\exp[-8\pi^2/g_{\rm YM}^2]$.
The coefficients $c_2$ and $c_3$ are positive ${\cal O}(1)$ quantities. 

We have treated the case of multiple instantons in the same manner, and found that the largest instanton among others dominates the amplitude. The instanton amplitude is again exponentially enhanced by the factor $\exp[c_2 \rho_{\rm largest} E]$ for the energy $E> c_3 \sqrt{\lambda}/\rho_{\rm largest}$.

We have performed a model analysis of the configuration of the string worldsheet. While even numerical study of the worldsheet is difficult due to the complexity of the boundary conditions (as we will discuss later in detail), a model analysis is enough for confirming the qualitative behavior of the scattering amplitudes found above. As we have shown in Sec.~\ref{sec:4}, a simple cut of the original Alday-Maldacena worldsheet by the instanton D3-brane exhibits the expected behavior anticipated in Sec.~\ref{sec:3}. In addition, we have studied a one-parameter family model of the worldsheet which respects a part of the D3-brane boundary condition. The obtained value of the action is lower than that of just the simple cut of the Alday-Maldacena configuration. These in total suggest that the topology change of the string worldsheet occurs when the instanton size is varied.

The findings above are consistent with the field theory picture, as follows. In the SYM, once the instanton is placed, the instanton amplitudes are made by gluons which hit and are scattered by the instantons. In the holographic and T-dual picture, these gluons are replaced by the string worldsheet in the momentum space. And indeed, our analysis shows that the instanton effect appears when the worldsheet ends partially on the instanton D3-brane, meaning that the string is scattered by the instanton D3-brane.

Even the instanton size dependence coincides with the field theory picture.
When the energy of the gluon increases, the null-polygon of the worldsheet boundary grows in size, and the worldsheet can extend deeper in the AdS, and reaches the instanton D3-brane closest to the AdS boundary at a certain critical energy. This corresponds to the fact in the SYM that the gluon wavelength gets shorter as the energy grows, and the gluons start to see the instanton once the gluon wavelength reaches the size of the largest instanton --- the critical energy.


Several discussions are in order.
\begin{itemize}
\item AI-assisted numerical method for generic minimal surface problem.

In this paper, we performed a model analysis for the shape of the worldsheet, because the numerical evaluation of the worldsheet is difficult due to the complexity of the boundary conditions. To resolve this issue, we may resort to a novel numerical technology of AI, which is the physics-informed neural network (PINN, \cite{raissi2019physics}). In our companion paper \cite{ours2}, we have a fully detailed analysis, and obtained the worldsheet configuration as shown in Fig.~\ref{fig:pinnours2}. 
Our model analysis with the worldsheet configuration of Fig.~\ref{fig:r_vs_y1_y2_065_080_088_110}
appears to be quite close in shape to the PINN solution of Fig.~\ref{fig:pinnours2}.

\begin{figure}[t]
    \centering
\includegraphics[width=9cm]{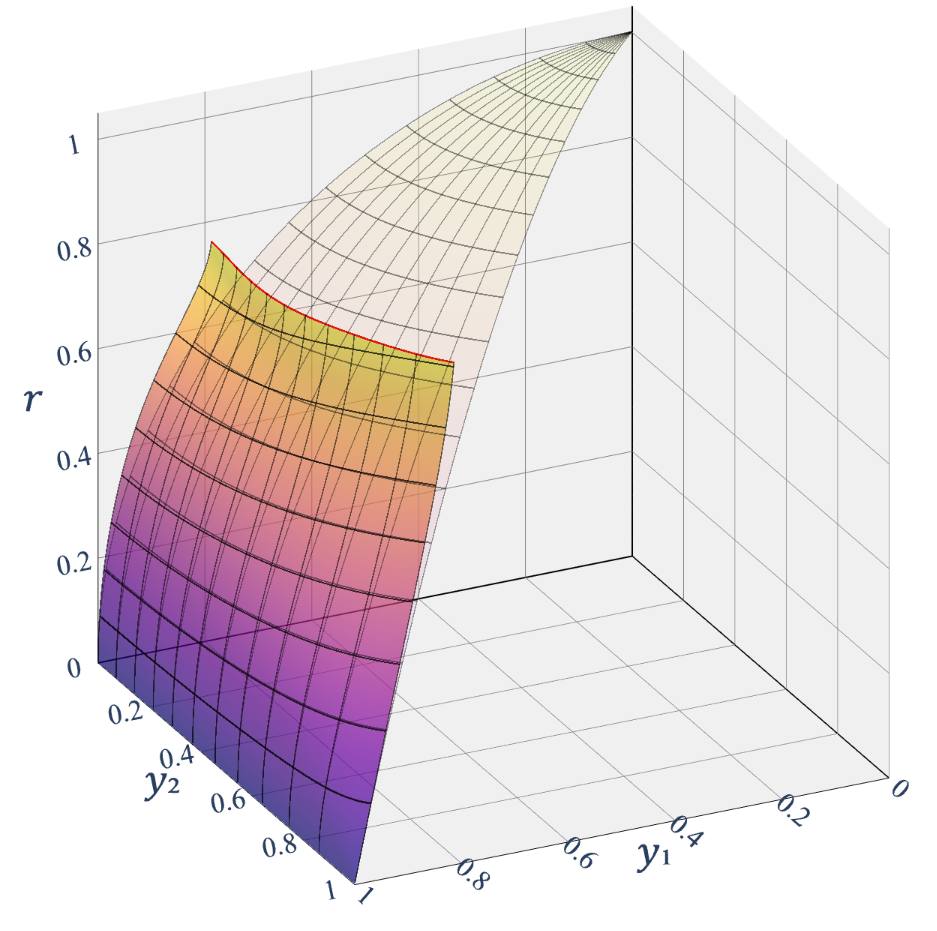}
    \caption{AI solution of the minimal surface, taken from our companion paper \cite{ours2}. Physics-informed neural networks are used to find this, and the technology and methods are described in detail in \cite{ours2}. The instanton D3-brane is put at $r=0.75$. Respecting the spacetime parity symmetry, only an eighth of the surface is shown. The Alday-Maldacena configuration is overlaid as transparent grid lines. }
\label{fig:pinnours2}
\end{figure}

The PINN could be widely applied to various situations in AdS/CFT.  
To evaluate nonlocal quantities holographically, we are forced to calculate the volume or the area of the minimal surfaces in the curved AdS-like geometries with various boundary conditions. The boundary conditions of the surface may be classified as follows: the condition at the AdS boundary, the condition at the black hole horizon, the condition at some other minimal surfaces.
Typically, the surface ends on a combination of these boundary conditions.
Consider, for example a flavor D-brane configuration. This is a minimal surface in typically the asymptotically AdS spacetimes. The flavor D-brane extends to the AdS boundary whose location corresponds to the quark mass. The flavor D-brane may fall into the black hole horizon in the deconfined phase, imposing another boundary condition for the D-brane configuration. If one considers the behavior of the quark, it is again a worldsheet minimal surface ending on the flavor D-brane and/or the black hole horizon.

Normally, the combination of various boundary conditions for anchoring the minimal surfaces to make sense in gravity theories makes the calculation of the volume difficult and intractable.\footnote{In fact, in the context of the bulk metric reconstruction via minimal surfaces (via entanglement \cite{Hammersley:2007ab}, Wilson loop \cite{Hashimoto:2020mrx}, or via complexity \cite{Hashimoto:2021umd}), multiple boundary conditions are often fatal.} 
Even numerically, imposing several boundary conditions at the same time provides intrinsic difficulty in performing the numerical evaluation.
When the minimal surface configuration has a symmetry and the Nambu-Goto equations of motion reduce to an ordinary differential equation (for example, for the case of circular Wilson loops), the situation is saved and one can solve it easily even with multiple boundary conditions. However, when there is no symmetry for the surface configuration, the governing equation becomes a set of partial differential equations. In such a situation, combined with various boundary conditions, even finding the topology of the surface may be difficult in general.

This reason has led us in our companion paper to the AI method: the physics-informed neural networks (PINNs) \cite{raissi2019physics}, which have been proven to be an efficient method to solve partial differential equations under a complicated set of boundary conditions. 
The PINN is nothing but an efficient machine learning method which seeks the minima of the loss function
$(\mbox{Equation of motion})^2  + (\mbox{Boundary condition})^2$.
At the minima, the configuration parametrized by a neural network will solve the equations of motion and at the same time various boundary conditions. The PINNs have been used in various situations ranging from industrial applications to purely theoretical considerations, and have been used for solving minimal surface configurations in general \cite{peng2021idrlnetphysicsinformedneuralnetwork, kabasi2023physics, kabasi2024gradient, ehebrecht2024pinntegrate, Mishra2025TPMS}. In the string theory context, the bulk reconstruction of the AdS/CFT correspondence was studied with the use of PINNs \cite{Ahn:2025tjp}. 

In our companion paper \cite{ours2}, we apply PINNs to solve worldsheet minimal surfaces in AdS with various boundary conditions. There, we check whether PINN works well or not for a popular example of the Wilson loops \cite{Maldacena:1998im}, and study the situation considered in this paper. We expect that the PINN can be used in a variety of gravity theories.

\item Instanton moduli integration.

Our study in this paper is on the gluon scattering amplitudes under just a fixed instanton background. To obtain the final expression for the instanton amplitudes, one needs to perform the instanton moduli integral;
\begin{align}
    {\cal A}_{\rm inst}
    = \int d^4X d\rho \frac{1}{\rho^5} {\cal A}_{\rm inst}(\rho)
    \label{eq:mi}
\end{align}
where the factor $1/\rho^5$ comes from the AdS invariant volume as the instanton moduli space metric is shown to be exactly AdS${}_5$ in the case of ${\cal N}=4$ SYM \cite{Dorey:1999pd, Bianchi:1998nk, Dorey:1998qh}.\footnote{A naive integral of the instanton location moduli $X$ diverges at the infrared, but one can argue that any contribution of the instanton should exist only when the instanton resides in the scattering region, which provides a certain IR cutoff.} 

In making the integral, an important fact is that the fermionic zero modes generally present in the background of instantons, responsible for the chiral anomaly. The amplitude \eqref{eq:mi} is nonzero only when the number of the fermionic zero modes matches, 
thus \eqref{eq:mi} needs to be supplemented with an appropriate number of gluino vertex insertions. In general, this produces additional final state legs which are soft fermions. 

The fermionic zero modes depend on $\rho$, therefore the integral has a further multiplicative factor of powers in $\rho$.
The integral over $\rho$ will be dominated at a certain saddle point $\rho=\rho_{\rm s}$.\footnote{It can also depend on the infrared cutoff of the instanton size. 
The physical manner of putting the infrared cutoff of the instanton size was discussed in \cite{Kharzeev:2001vs}, and in realistic use of
the instanton amplitudes is expected to be infrared safe.} Then the rough estimate of the instanton amplitude is
\begin{align}
    {\cal A}_{\rm inst}
    \propto \exp\left[-\frac{8\pi^2}{g_{\rm YM}^2}\right]
    \exp\left[\tilde{c}E \rho_{\rm s}\right]
\end{align}
with some positive ${\cal O}(1)$ constant $\tilde{c}$.
This grows rapidly in gluon energy. The physical consequence of this behavior calls for more study.\footnote{In fact, in QCD the renormalization group and the asymptotic freedom induce further modification of the $\rho$ integral, and the high energy behavior could be drastically different from that of the conformal SYM. See \cite{Khoze:1991mx,mueller1991first,mueller1991higher} for early references of the instanton effects in gluon scattering (see also \cite{Khoze:2019jta} for the relevance to collider experiments).}

It would be interesting to compare our findings with the issue of thermalization of ${\cal N}=4$ SYM. In \cite{Kharzeev:2009pa}, it was shown that the total cross section of gluon collision grows in energy due to the D-instanton back reaction, and the multi-instanton process contributes to the thermalization of ${\cal N}=4$ SYM. On the other hand, our finding suggests the enhancement even without the backreaction. It would be interesting if our findings can shed some light on the issue of thermalization in ${\cal N}=4$ SYM.

\end{itemize}

In addition to these discussions, as we have emphasized in Sec.~\ref{sec:3}, the minimal surface appears on many occasions in AdS/CFT,
we expect that this may lead to some new duality relation in string theory. Our gluon scattering amplitudes are a worldsheet with a cylindrical topology with the topology change, so the typical behavior of it is quite similar to the other minimal surfaces listed in Table \ref{tab:AdS_CFT_dictionary}. It is possible that there exists a deeper connection, for example along the line of \cite{Alday:2008yw}. Further study is necessary to see the unified view of minimal surfaces in curved geometry, with the help of an AI-assisted numerical method.


\section*{Acknowledgments}

K.~H.~would like to thank V.~Khoze for valuable discussions.
The work of K.~H.~was supported in part by JSPS KAKENHI Grant No.~JP22H01217, JP22H05111 and JP22H05115.
The work of N.~T.~was supported in part by JSPS KAKENHI Grant No.~JP21H05189, JP22H05111 and 25K07282.


\appendix
\section{The Minimal Surface in Global Coordinates}
\label{app:1}

In this appendix, we provide a more detailed description of the minimal surface (\ref{eq:1cusp}). As noted in \cite{Alday:2007hr}, this surface appears to have a single cusp but in fact possesses four cusps located on the AdS boundary. This apparent discrepancy can be clarified by switching to global coordinates, as three of the cusps lie at infinity in the Poincaré patch and are thus not visible. By employing AdS$_5$ isometries—which correspond to conformal transformations on the AdS boundary—we can map all four cusps into a single Poincar\'e patch.

To describe this coordinate transformation, it is convenient to work in the embedding coordinates $\{Y^M\}$ ($M = -1, 0, \ldots, 4$), since AdS$_5$ isometries act as simple $SO(2,4)$ rotations in this representation. The embedding coordinates relate to the Poincaré and global coordinates as follows:
\begin{align}
    &(\mathrm{Embedding})&&\ \ \ \ (\mathrm{Poincare})&&\ \ \ \ \ (\mathrm{Global})\nonumber \\
    &Y^{-1}&&=\frac{1}{2r}\left(R^2+r^2+y^\mu y_\mu\right)&&=R\frac{\cos\tau}{\cos\varrho}\nonumber \\
    &Y^0&&=\frac{1}{r}Ry^0&&=R\frac{\sin\tau}{\cos\varrho}\nonumber \\
    &Y^1&&=\frac{1}{r}Ry^1&&=R\tan\varrho\sin\theta_3\sin\theta_2\sin\theta_1\nonumber \\
    &Y^2&&=\frac{1}{r}Ry^2&&=R\tan\varrho\sin\theta_3\sin\theta_2\cos\theta_1\nonumber \\
    &Y^3&&=\frac{1}{r}Ry^3&&=R\tan\varrho\sin\theta_3\cos\theta_2\nonumber \\
    &Y^4&&=\frac{1}{2r}\left(R^2-r^2-y^\mu y_\mu\right)&&=R\tan\varrho\cos\theta_3 \,.
    \label{eq:AdS5_coord_trans}
\end{align}
Here, $\tau\in\mathbb{R}$, $\varrho\in[0,\frac{\pi}{2}]$, $\theta_1 \in[0,2\pi]$, and $\theta_2, \theta_3 \in[0,\pi]$ are global coordinates, while $r>0$ and $y^\mu\in\mathbb{R}$ are Poincaré coordinates. In embedding coordinates, AdS$_5$ is described as the hyperboloid $-Y_{-1}^2 - Y_0^2 + Y_1^2 + \cdots + Y_4^2 = -R^2$ in $\mathbb{R}^{2,4}$, where $R$ denotes the AdS radius. We leave $R$ arbitrary throughout this appendix.

In Poincaré coordinates, the cusp solution (\ref{eq:1cusp}) is given by
\begin{align}
    y^0&=Re^{\sigma_2}\cosh\sigma_1 \ ,&&  y^1=Re^{\sigma_2}\sinh\sigma_1 \nonumber\\ 
    r&=Re^{\sigma_2}\sqrt{2} \ ,   && y^2=y^3=0 \ \label{eq:1cusp_exact_in_Poin},
\end{align}
where $\sigma_1$ and $\sigma_2$ are worldsheet coordinates. Using the relations in (\ref{eq:AdS5_coord_trans}), this surface can be expressed in embedding and global coordinates as:
\begin{align}
    (\mathrm{Embedding}):\left\{\begin{array}{lll}
         Y^{-1}=\frac{R}{\sqrt{2}}\cosh\sigma_2\ , \ \ \
         & Y^1=\frac{R}{\sqrt{2}}\sinh\sigma_1 \ , \ 
         &Y^2=0 , \\
         Y^0\ \ =\frac{R}{\sqrt{2}}\cosh\sigma_1\ ,
         &Y^4=\frac{R}{\sqrt{2}}\sinh\sigma_2 \ ,
         & Y^3=0
    \end{array}\right. \label{eq:1cusp_exact_in_emb}\\
    (\mathrm{Global}):\left\{\begin{array}{ll}
         \sin\tau=\frac{\cosh\sigma_1}{\sqrt{\cosh^2\sigma_1+\cosh^2\sigma_2}}\ ,
         &\sin\varrho=\sqrt{\frac{\sinh^2\sigma_1+\sinh^2\sigma_2}{\cosh^2\sigma_1+\cosh^2\sigma_2}} \ ,\\
         \sin\theta_3=\frac{|\sinh\sigma_1|}{\sqrt{\sinh^2\sigma_1+\sinh^2\sigma_2}}\ ,& \sin\theta_1=\mathrm{sign}(\sigma_1) \ , \ \ \ \theta_2=\frac{\pi}{2} 
    \end{array}\right.  \nonumber
\end{align}
The AdS boundary in global coordinates, $\varrho \rightarrow \frac{\pi}{2}$, corresponds to the limits $\sigma_1 \rightarrow \pm\infty$ or $\sigma_2 \rightarrow \pm\infty$ on the worldsheet. The cusp at the origin in Poincaré coordinates (\ref{eq:1cusp_exact_in_Poin}) corresponds to the limit $(\sigma_2 \rightarrow -\infty,\ \sigma_1:\ \mathrm{finite})$, while the remaining three cusps correspond to $(\sigma_2 \rightarrow +\infty,\ \sigma_1:\ \mathrm{finite})$ and $(\sigma_1 \rightarrow \pm\infty,\ \sigma_2:\ \mathrm{finite})$, all of which lie on the boundary of the Poincaré patch.

We can map the entire surface (\ref{eq:1cusp_exact_in_Poin}) into a finite region of the Poincaré patch by applying an appropriate $SO(2,4)$ transformation. To reproduce the configuration in \cite{Alday:2007hr}, we perform the following three rotations:
\begin{align*}
  &Y\longrightarrow 
  O_{1,2}(-\pi/4)\ O_{2,4}(\pi/2)\ O_{-1,0}(\pi/4) \ Y
\end{align*}
Here, $O_{M,N}(\phi)$ denotes a rotation in the $(Y^M,Y^N)$ plane by angle $\phi$ in embedding coordinates:
\begin{align*}
    &O_{-1,0}\left(-\frac{\pi}{4}\right)=
  \left(\begin{array}{c|c}
      \begin{array}{cc}
        1/\sqrt{2}&1/\sqrt{2}\\
        -1/\sqrt{2}&1/\sqrt{2} 
      \end{array} & \\
      \hline
      & I_4 
  \end{array}\right)
  &&O_{24}\left(\frac{\pi}{2}\right)=
  \left(\begin{array}{c|c}
      I_2& \\
      \hline
      &\begin{array}{cccc}
        1& & & \\
        & &  & -1 \\
        & & 1&  \\
        &1& &  \\ 
      \end{array}
  \end{array}\right)\\
  &O_{1,2}\left(\frac{\pi}{4}\right)=
  \left(\begin{array}{c|c|c}
      I_2 & & \\
        \hline
      & \begin{array}{cc}
        1/\sqrt{2}&-1/\sqrt{2}\\
        1/\sqrt{2}&1/\sqrt{2} 
      \end{array} & \\
      \hline
      & & I_2 
  \end{array}\right) \,.
\end{align*}
Applying these transformations, we obtain
\begin{align}
    Y^{-1}=\frac R2(\cosh\sigma_1+\cosh\sigma_2) \ , && Y^{0}=\frac R2(\cosh\sigma_1-\cosh\sigma_2)
    \nonumber\\
    Y^{1}=\frac R2(\sinh\sigma_1+\sinh\sigma_2)\ , && Y^{2}=\frac R2(\sinh\sigma_1-\sinh\sigma_2)
    \ ,&&
    Y^3=0=Y^4 \,. \label{eq:4cusps_exact_in_Poin}
\end{align}
Transforming back to Poincaré coordinates, we find:
\begin{align}
    &y^0=R\frac{\cosh\sigma_1-\cosh\sigma_2}{\cosh\sigma_1+\cosh\sigma_2}\ ,&&
    y^1=R\frac{\sinh\sigma_1+\sinh\sigma_2}{\cosh\sigma_1+\cosh\sigma_2} \ , 
    &y^2=R\frac{\sinh\sigma_1-\sinh\sigma_2}{\cosh\sigma_1+\cosh\sigma_2}\\
    &y^3=0 \ , &&r=R\frac{2}{\cosh\sigma_1+\cosh\sigma_2}
\end{align}
which reduces to the original expression of the minimal surface with a square boundary \cite{Alday:2007hr},
\begin{align*}
    &y^0=y^1y^2  \ , &&r=\sqrt{(1-(y^1)^2)(1-(y^2)^2)} \ , &&y^3=0 \,,
\end{align*}
upon setting $R = 1$.

\section{Discussion on the model parameter space}
\label{app:2}

In Sec.~\ref{subsec:model2}, we have defined a three-parameter model and presented a one-parameter slice which is scanned. In this appendix, we study the total three-dimensional parameter space and discuss its critical issue: the model may not allow a local minimum of the action, at least within our search of the total three-dimensional parameter space.

The result of the one-dimensional slice in the parameter space is shown in Fig.~\ref{fig:SNG_r110_t065}. One notices that in the figure, when we vary $\tau_2$ too, the minimum action value $S_{>\tau_0}$ decreases as we take smaller $\tau_2$. 

\begin{figure}[b]
\centering
   \includegraphics[height=8cm]{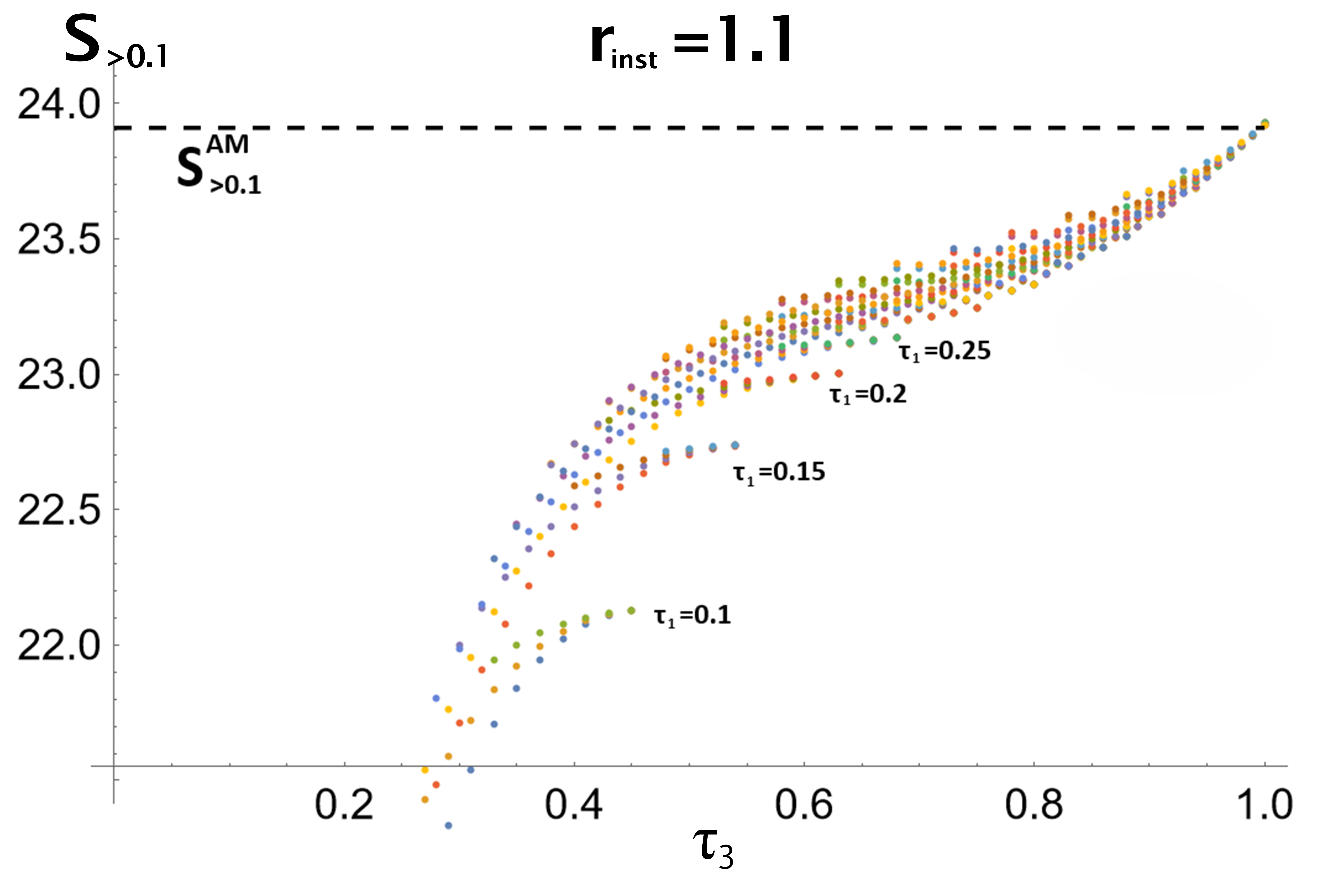}
    \caption{Regularized worldsheet area $S_{>0.1}$ of $r$ reaching the instanton D3-brane placed at $r_{\rm inst}=1.1$. Compared to Fig.~\ref{fig:SNG_r110_t065}, in this plot the parameter $\tau_1$ is also varied from $\tau_1=0.1$ to $\tau_1=0.8$. In each $\tau_1$, reducing the values of $\tau_2$ and $\tau_3$ makes $S$ smaller. }
    \label{fig:r110}
\end{figure}

Furthermore, when we take $\tau_1$ smaller, the lower these plots shift (see Fig.~\ref{fig:r110}). This means that, among our ansatz, the worldsheet is forced to have a wider hole, eventually becoming the trivial null surface mentioned before. In other words, there is no parameter set $(\tau_1,\tau_2,\tau_3)$ which gives a local minimum (or maximum) of $S_{>0.1}$. All of $\tau_1,\tau_2,\tau_3$ go to zero (or one), giving zero Nambu-Goto area (or the same value as $S^{AM}$). 

\begin{figure}[h]
  \centering
  \begin{minipage}[b]{0.49\columnwidth}
    \centering
    \includegraphics[width=0.9\columnwidth]{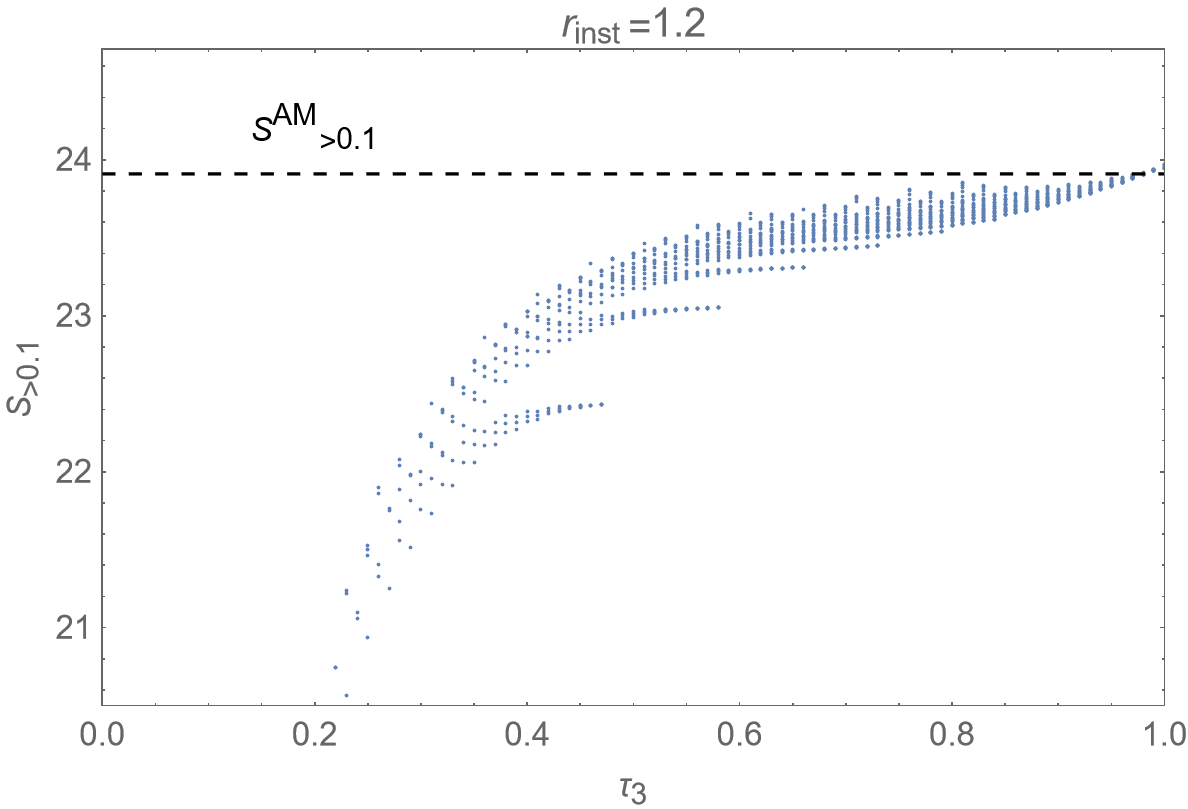}
  \end{minipage}
  \begin{minipage}[b]{0.49\columnwidth}
    \centering
    \includegraphics[width=0.9\columnwidth]{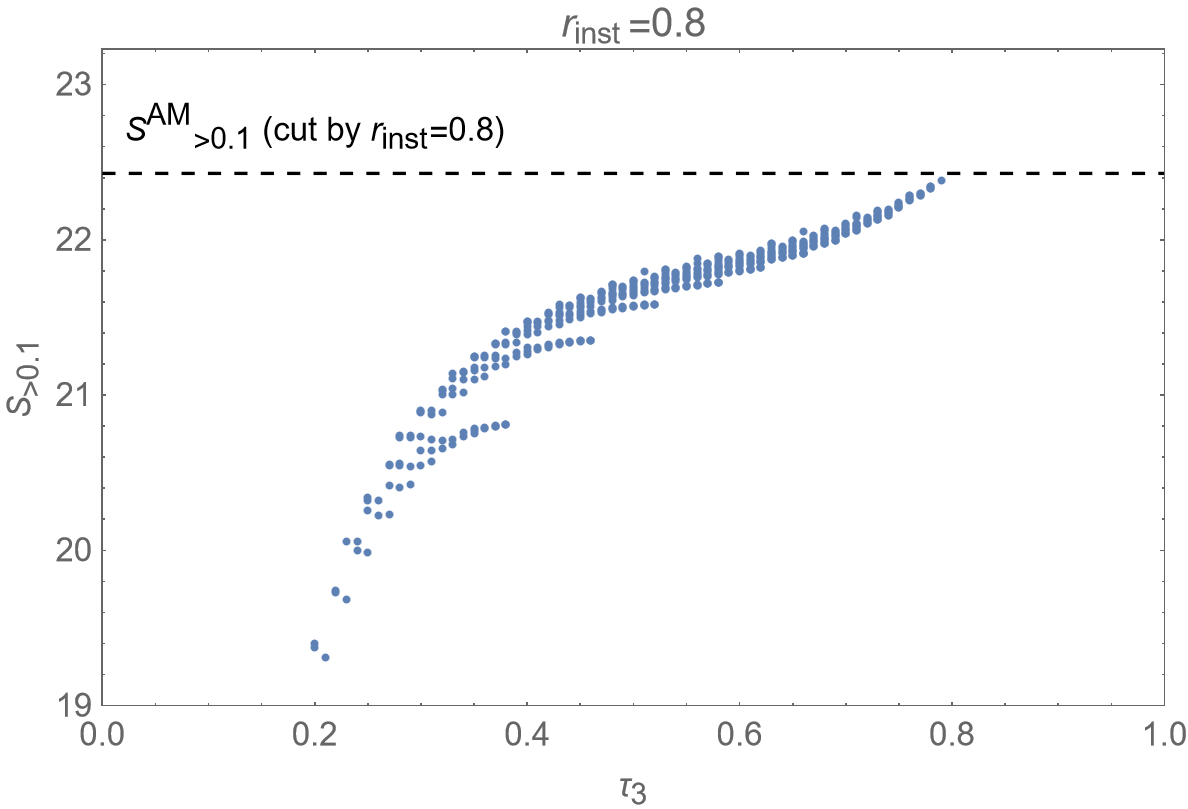}
  \end{minipage}
  \caption{The Nambu-Goto area $S_{>0.1}$ of the configurations with various parameters $(\tau_1,\tau_2,\tau_3)$, for $r_{\rm inst}=1.2$ (Left) and $r_{\rm inst}=0.8$ (Right). When $\tau_1=\tau_2=\tau_3=1$ (or $=\sqrt{r_{\rm inst}}$ if $r_{\rm inst}<1$), the configuration coincides with the original one (or cut by $r=r_{\rm inst}$), giving the same value as $S_{>0.1}^{\rm AM}$. On the other hand, taking $\tau_1=\tau_2=\tau_3=0$ leads to the trivial null surface, giving zero Nambu-Goto area.}  
  \label{fig:SNG_list}
\end{figure}

The D3-brane distance $r_{\rm inst}$ is varied from $0.8$ to $1.2$ in $0.1$ increments. In each case, within our numerical scan, we find that the optimization of $\tau_1,\tau_2,\tau_3$ leads to the trivial null surface. (See Fig.~\ref{fig:SNG_list} )

Of course, this observation is obtained within our model ansatz and within our scan. 
Our model ansatz has a critical limitation, and the possibility remains that there are some other parameterizations of the worldsheet which allow a local minimum.\footnote{Note that in our model we do not modify the $y_0$ configuration and just use the cut of Alday Maldacena configuration, so the boundary condition for $y_0$ has not been satisfied on the D3-brane. This could be a critical reason for the absence of the local minimum.}
However, we stress that the model analysis works to roughly evaluate the value of the action with the topology change, as we find that the action value with our ansatz is of the same order as that of Alday-Maldacena. This suggests that in fact the topology change can occur.

We propose several reasons for this monotonicity. First, to satisfy the boundary condition imposed on the AdS boundary, the worldsheet gets more lightlike when approaching the boundary. This nature of the boundary attracts the surface elements and the endline on the D3-brane to the trivial null solution. Although there is an effect counteracting this nature to shrink the hole in the way that expanding the cylinder radius $\sqrt{y_1^2+y_2^2}$ costs more surface element in the $(y_1,y_2,r)$ space, the former effect of the null directions seems to overcome the latter. 

\begin{figure}[h]
\centering
   \includegraphics[height=8cm]{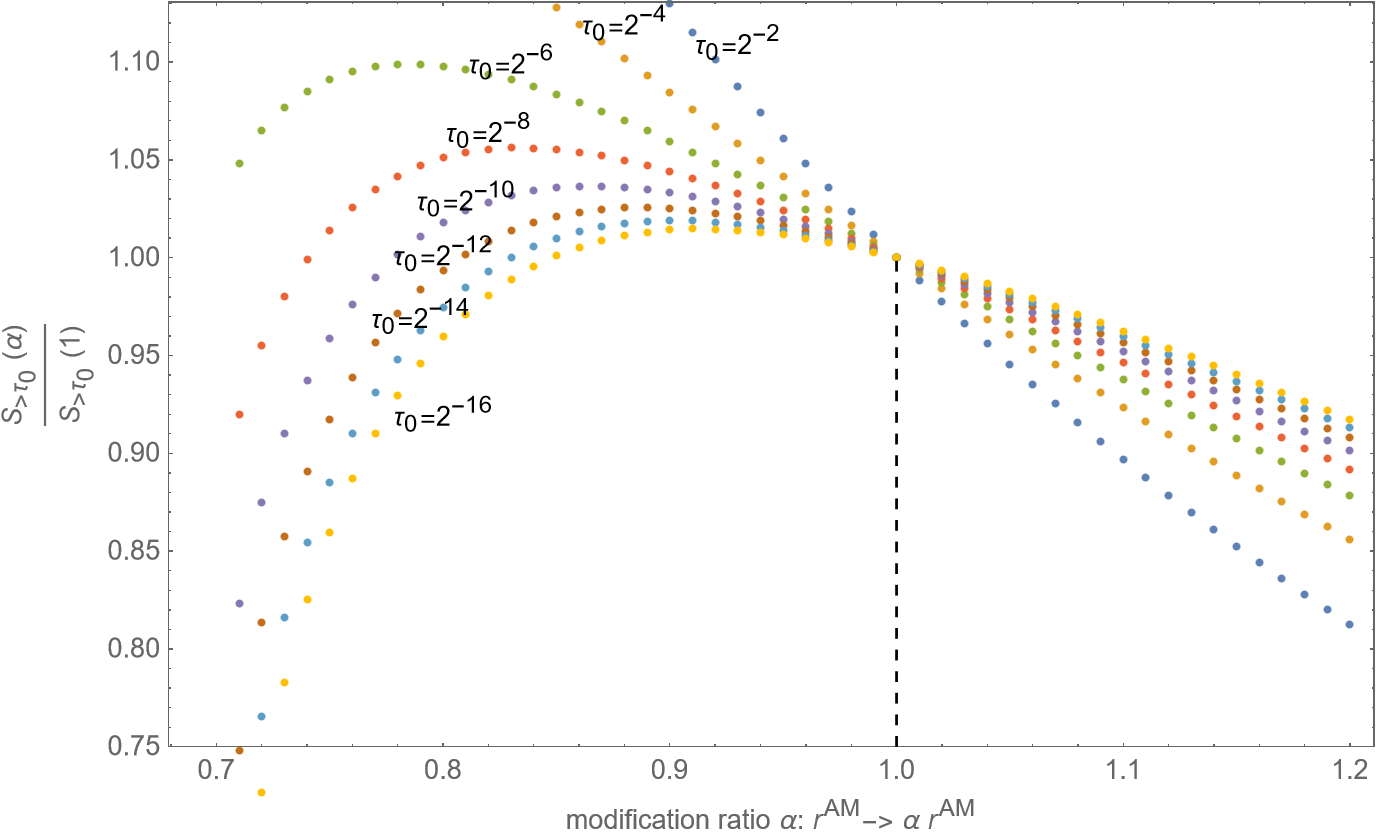}
    \caption{The $\alpha$ dependence of $S_{>\tau_0}(\alpha)$ normalized by $S_{>\tau_0}(\alpha=1)$. The cutoff parameter $\tau_0$ is varied exponentially from $2^{-2}$ to $2^{-16}$, corresponding to each color. The maximum point seems to approach $\alpha =1$ . }
    \label{fig:limit}
\end{figure}

Another possible reason for the monotonicity is that the original solution $r^{\rm AM},y_0^{\rm AM}$ may not be a minimum of the action but a saddle point or even a maximum of the action, so the configurations constructed by modifying the original solution tend to roll down the slope in a similar configuration space. In fact, considering a one-parameter group of the worldsheet configurations parameterized by $\alpha$:
 \begin{align*}
     &r(y_1,y_2)=\alpha \  r^{\rm AM}(y_1,y_2) && y_0(y_1,y_2)=y_0^{AM}(y_1,y_2) \ 
 \end{align*}
 and its Nambu-Goto area
 \begin{align*}
     S_{>\tau_0}(\alpha) = \iint_{\tau(y_1,y_2)>\tau_0} dy_1 dy_2 \frac{\sqrt{1-(1-\alpha^2)(y_1^2+y_2^2)}}{\alpha^2\ \tau(y_1,y_2)^2} \ ,
 \end{align*}
the original solution corresponding $\alpha=1$ turns out to be a maximum point of $S(\alpha)$. More precisely, the maximum point $\alpha_{\rm max}$ depends on the cutoff parameter $\tau_0$, but in the cutoff zero limit $\tau_0\rightarrow0$,  $\alpha_{\rm max}$ approaches one. This limiting behavior is shown in Fig.~\ref{fig:limit}. 

To fix this issue, we may have to consider more general worldsheet configurations, removing the restriction that the worldsheet should asymptotically coincide with the no-hole solution $(r,y_0)=(r^{\rm AM},y_0^{\rm AM})$. 

\bibliographystyle{paper1}
\bibliography{For_paper1}

%



\end{document}